\documentclass[twocolumn]{aastex631}

\usepackage[utf8]{inputenc}
\usepackage{comment}

\newcommand{\angstrom}{\text{\normalfont\AA}}

\defcitealias{Merlin2022}{Paper II}
\defcitealias{Castellano2022}{Paper III}
\defcitealias{Santini2022}{Paper XI}
\defcitealias{Treu2022b}{Paper XII}
\defcitealias{Yang2022}{Paper V}

\accepted{September 21, 2022}

\shorttitle{Z-dropout properties from JWST imaging}
\shortauthors{Leethochawalit et al.}

\usepackage{amsmath}
\begin{document}

\title{Early results from GLASS-JWST. X: Rest-frame UV-optical properties of galaxies at $7<z<9$}


\correspondingauthor{Nicha Leethochawalit}
\email{nicha.leethochawalit@unimelb.edu.au}

\author[0000-0003-4570-3159]{N. Leethochawalit}
\affiliation{School of Physics, University of Melbourne, Parkville 3010, VIC, Australia}
\affiliation{ARC Centre of Excellence for All Sky Astrophysics in 3 Dimensions (ASTRO 3D), Australia}
\affiliation{National Astronomical Research Institute of Thailand (NARIT), Mae Rim, Chiang Mai, 50180, Thailand}

\author[0000-0001-9391-305X]{M. Trenti}
\affiliation{School of Physics, University of Melbourne, Parkville 3010, VIC, Australia}
\affiliation{ARC Centre of Excellence for All Sky Astrophysics in 3 Dimensions (ASTRO 3D), Australia}

\author[0000-0002-9334-8705]{P.~Santini}
\affiliation{INAF Osservatorio Astronomico di Roma, Via Frascati 33, 00078 Monteporzio Catone, Rome, Italy}

\author[0000-0002-8434-880X]{L.~Yang}
\affiliation{Kavli Institute for the Physics and Mathematics of the Universe, The University of Tokyo, Kashiwa, Japan 277-8583}

\author[0000-0001-6870-8900]{E.~Merlin}
\affiliation{INAF Osservatorio Astronomico di Roma, Via Frascati 33, 00078 Monteporzio Catone, Rome, Italy}

\author[0000-0001-9875-8263]{M.~Castellano}
\affiliation{INAF Osservatorio Astronomico di Roma, Via Frascati 33, 00078 Monteporzio Catone, Rome, Italy}

\author[0000-0003-3820-2823]{A. Fontana}
\affiliation{INAF Osservatorio Astronomico di Roma, Via Frascati 33, 00078 Monteporzio Catone, Rome, Italy}

\author[0000-0002-8460-0390]{T. Treu}
\affiliation{Department of Physics and Astronomy, University of California, Los Angeles, 430 Portola Plaza, Los Angeles, CA 90095, USA}

\author[0000-0002-3407-1785]{C. Mason}
\affiliation{Cosmic Dawn Center (DAWN), Denmark}
\affiliation{Niels Bohr Institute, University of Copenhagen, Jagtvej 128, DK-2200 Copenhagen N, Denmark}

\author[0000-0002-3254-9044]{K. Glazebrook}
\affiliation{Centre for Astrophysics and Supercomputing, Swinburne University of Technology, PO Box 218, Hawthorn, VIC 3122, Australia}

\author[0000-0001-5860-3419]{T. Jones}
\affiliation{Department of Physics and Astronomy, University of California Davis, 1 Shields Avenue, Davis, CA 95616, USA}

\author[0000-0003-0980-1499]{B. Vulcani}
\affiliation{INAF Osservatorio Astronomico di Roma, Via Frascati 33, 00078 Monteporzio Catone, Rome, Italy}

\author[0000-0003-2804-0648 ]{T.~Nanayakkara}
\affiliation{Centre for Astrophysics and Supercomputing, Swinburne University of Technology, PO Box 218, Hawthorn, VIC 3122, Australia}
\affiliation{ARC Centre of Excellence for All Sky Astrophysics in 3 Dimensions (ASTRO 3D), Australia}

\author[0000-0001-9002-3502]{D.~Marchesini}
\affiliation{Department of Physics and Astronomy, Tufts University, 574 Boston Ave., Medford, MA 02155, USA}
\author[0000-0002-9572-7813]{S.~Mascia}
\affiliation{INAF Osservatorio Astronomico di Roma, Via Frascati 33, 00078 Monteporzio Catone, Rome, Italy}

\author[0000-0002-8512-1404]{T. Morishita}
\affiliation{Infrared Processing and Analysis Center, Caltech, 1200 E. California Blvd., Pasadena, CA 91125, USA}

\author[0000-0002-4140-1367]{G. Roberts-Borsani}
\affiliation{Department of Physics and Astronomy, University of California, Los Angeles, 430 Portola Plaza, Los Angeles, CA 90095, USA}

\author{A.~Bonchi}
\affiliation{INAF Osservatorio Astronomico di Roma, Via Frascati 33, 00078 Monteporzio Catone, Rome, Italy}
\affiliation{ASI-Space Science Data Center,  Via del Politecnico, I-00133 Roma, Italy}

\author{D.~Paris}
\affiliation{INAF Osservatorio Astronomico di Roma, Via Frascati 33, 00078 Monteporzio Catone, Rome, Italy}

\author[0000-0003-4109-304X]{K.~Boyett}
\affiliation{School of Physics, University of Melbourne, Parkville 3010, VIC, Australia}
\affiliation{ARC Centre of Excellence for All Sky Astrophysics in 3 Dimensions (ASTRO 3D), Australia}

\author[0000-0002-6338-7295]{V.~Strait}
\affiliation{Cosmic Dawn Center (DAWN), Denmark}

\author[0000-0003-2536-1614]{A. Calabr\`o}
\affiliation{INAF Osservatorio Astronomico di Roma, Via Frascati 33, 00078 Monteporzio Catone, Rome, Italy}

\author[0000-0001-8940-6768 ]{L. Pentericci}
\affiliation{INAF Osservatorio Astronomico di Roma, Via Frascati 33, 00078 Monteporzio Catone, Rome, Italy}

\author[0000-0001-5984-0395]{M. Bradac}
\affiliation{University of Ljubljana, Department of Mathematics and Physics, Jadranska ulica 19, SI-1000 Ljubljana, Slovenia}
\affiliation{Department of Physics and Astronomy, University of California Davis, 1 Shields Avenue, Davis, CA 95616, USA}

\author[0000-0002-9373-3865]{X. Wang}
\affil{Infrared Processing and Analysis Center, Caltech, 1200 E. California Blvd., Pasadena, CA 91125, USA}

\author[0000-0002-9136-8876]{C.~Scarlata}\affiliation{
School of Physics and Astronomy, University of Minnesota, Minneapolis, MN, 55455, USA}

\begin{abstract}

We present the first James Webb Space Telescope/NIRCam-led determination of $7<z<9$ galaxy properties based on broadband imaging from 0.8 to 5~$\mathrm{\mu m}$ as part of the GLASS-JWST Early Release Science program. This is the deepest dataset acquired at these wavelengths to date, with an angular resolution $\lesssim0.14$ arcsec. We robustly identify 13 galaxies with $S/N\gtrsim8$ in F444W from 8 arcmin$^2$ of data at $m_{AB}\leq 28$ from a combination of dropout and photometric redshift selection. From simulated data modeling, we estimate the dropout sample purity to be $\gtrsim90\%$. We find that the number density of these F444W-selected sources is broadly consistent with expectations from the UV luminosity function determined from Hubble Space Telescope data. We characterize galaxy physical properties using a Bayesian Spectral Energy Distribution fitting method, finding median stellar mass $10^{8.5}M_\odot$ and age 140 Myr, indicating they started ionizing their surroundings at redshift $z>9.5$. Their star formation main sequence is consistent with predictions from simulations. Lastly, we introduce an analytical framework to constrain main-sequence evolution at $z>7$ based on galaxy ages and basic assumptions, through which we find results consistent with expectations from cosmological simulations. While this work only gives a glimpse of the properties of typical galaxies that are thought to drive the reionization of the universe, it clearly shows the potential of JWST to unveil unprecedented details on galaxy formation in the first billion years.

\end{abstract}

\keywords{galaxies: evolution – galaxies: fundamental parameters – galaxies: high-redshift}

\section{Introduction} \label{sec:intro}
Redshift $z\sim8-7$ is likely the epoch when the universe undergoes a rapid phase transition. The intergalactic medium neutral fraction $\overline{x}_\textsc{hi}$ quickly declines from $\overline{x}_\textsc{hi}\sim1$ to $\sim0$ with decreasing redshift \citep[e.g.,][]{Treu2013,McGreer2015,Greig2017,Banados2018,Mason2018}. For a full understanding of this complex reionisation process, it is fundamental to characterise the physical properties of the sources, especially galaxies around and below the characteristic luminosity $L_*$, as those are thought to contribute the most to the photon budget due to their high abundance \citep[e.g.,][]{Robertson2015}. Key topics to study include, but are not limited to, their star formation rates, underlying stellar population properties such as mass, ages and metallicities, as well as the morphology and dust content.

Spectral energy distribution (SED) fitting is a crucial process to derive galaxy physical parameters \citep[e.g.,][]{Maraston2010,rb21}. So far, the combination of Hubble and IRAC/Spitzer has been the primary tool for the community to determine  SED properties of $z\gtrsim7$ galaxies. However, IRAC photometry is limited in terms of both depth and spatial resolution, which can lead to systematic uncertainties from deblending procedures \citep{Merlin2015,Tacchella2022}. The deepest data set on Spitzer/IRAC have $5\sigma$ depths of $\sim26-27$ mag in the 4.5 $\mu$m band \citep[e.g.][]{Labbe2015ApJS,Stefanon2022b}, which limit the current  measurements to bright massive galaxies (stellar masses $>10^9M_\odot$, e.g.  \citealt{Roberts-Borsani2022}). For lower mass galaxies, the measurements are mainly based on stacked SEDs \citep[e.g.][]{Stefanon2022a_stack} or a handful of gravitationally lensed sources \citep[e.g.][]{Castellano2016,Atek2018,Bradac2019,strait21}. 

Moreover, given Spitzer's limited angular resolution, most of the existing $z\gtrsim7$ galaxies are selected via detection in rest-frame UV bands either from Hubble Space Telescope (HST) images (e.g. \citealt{Labbe2010,Roberts-Borsani2016}), or from ground-based surveys \citep[e.g.,][]{Whitler2022}. Interestingly, ALMA has recently been playing an important role in high-z studies, yielding unprecedented observations of galaxies that are dark in HST bands \citep{Franco2018,Shu2022}. These results show that the selection of high-$z$ galaxies based on rest-frame UV are partially biased and that redder wavelengths can potentially uncover an undiscovered population. Selection of $z>7$ galaxies from rest-frame optical band (i.e. $\sim 4~\mathrm{\mu m}$ observer-frame) has not been possible so far, but the situation has now changed with the availability of James Webb Space Telescope (JWST hereafter) NIRCam data.

In this letter, we present F444W-selected F090W-dropout candidates ($7\lesssim z \lesssim 9$) in the NIRCam parallel pointing from the JWST GLASS Early Release Science Program \citep{Treu2022}, the deepest image at these wavelengths obtained so far from the new observatory. This is effectively approximately a rest-frame $V$-band selection. We report the physical properties of the identified candidates, such as luminosities, stellar masses, star-formation rates (SFRs), and ages. Further, we use the galaxy properties to constrain the star-formation main sequence at this redshift and above. With the sensitivity of JWST and depth of the ERS program, we investigate galaxies with $10^8\lesssim M_*/M_\odot\lesssim10^9$ (F444W$\lesssim 28$ mag), approaching a mass range that likely includes many of the sources that are responsible for reionization \citep[e.g.,][]{Alvarez2012}. 

We quote magnitudes in AB system \citep{Oke1983} and adopt the standard cosmology with $\Omega_{\rm m}=0.3$, $\Omega_{\Lambda}=0.7$ and H$_0$=70 km s$^{-1}$ Mpc$^{-1}$.

\section{Data and Sample Selection} 
We use the photometric catalog of the first NIRCam parallel observations of the GLASS-JWST ERS program obtained on June 28-29, 2022 \citep[Paper II][]{Merlin2022}. The observation was acquired in parallel to the primary NIRISS observation of cluster Abell 2744. The field is sufficiently far to avoid significant gravitational-lensing magnification (approximately one virial radius away from the cluster core). Although magnification at the $\sim 30
\%$ level is still expected based on weak lensing studies \citep{Medezinski2016}, we assume that all galaxies in this work have magnification $\mu$=1. The dataset includes seven wide filters: F090W, F115W, F150W, F200W, F277W, F356W, and F444W. Their $5\sigma$ depths in a $0.30\arcsec$-diameter aperture are in the range $29-29.5$ AB mag.
We refer readers to \citetalias{Merlin2022} for the detailed description of the image reduction, sensitivity, and photometric catalog construction. The final images are PSF-matched to the F444W band (coarsest resolution with FWHM $\approx0.14\arcsec$). Objects are detected with SourceExtractor \citep{Bertin1996} using F444W as the detection image. I.e. for $z\sim7-8$, the galaxies are detected based on the rest-frame V band. In this work, the colors refer to magnitudes measured in circular apertures of size $0.30\arcsec$ in diameter ($\sim2$ FWHM). The total F444W fluxes are the Kron fluxes measured with \texttt{A-PHOT} \citep{Merlin2019}, taken directly from the catalog in \citetalias{Merlin2022}. For other bands, the total fluxes are $0.30\arcsec$-diameter aperture fluxes scaled with the ratios between the total fluxes and the $0.30\arcsec$-diameter aperture fluxes in the F444W band. Our aperture size choise is motivated so that it captures most of the fluxes from the galaxies at these redshifts while minimising risk of contamination from other nearby sources. As measured in \citet[Paper V]{Yang2022}, our final galaxy candidates have median effective radius of 0.54 kpc in the F444W band, which corresponds to 0.10\arcsec at $z=8$). We note if we were to use 0.45\arcsec-diameter aperture fluxes to scale total flux calculations, our results would remain substantially unchanged within the limits of photometric scatter uncertainties.

\subsection{Color and SNR selection criteria}\label{sec:selection_criteria}
We optimize the selection criteria by following the framework described in \citet{Hainline2020} based on the JAdes extraGalactic Ultradeep Artificial Realizations (\texttt{JAGUAR}) mock catalog \citep{Williams2018}. We aim to obtain a high-purity F090W dropout sample, which led to the following criteria:
\begin{equation}
    \begin{aligned}
        &\textrm{F090W}-\textrm{F115W}>0.75\\
        &\textrm{F150W}-\textrm{F200W} < 0.4\\
        &\textrm{F090W}-\textrm{F115W} > 1.5\times(\textrm{F150W}-\textrm{F200W})+1\\
        &\textrm{SNR(F090W)}<2
.0    \end{aligned}
\end{equation} 
We also require that the signal-to-noise ratio (SNR) is $>8$ in the F444W band and $>2$ in the F115W, F150W, and F200W bands. We note that if sources satisfy the selection criteria for F115W dropouts ($z>9$) as presented in \citet[Paper III]{Castellano2022}, they are included in that sample instead.

A four-band color selection is recommended by \citet{Hainline2020} as it yields higher completeness and accuracy than a three-band color selection. Generally, the non-detection criteria are only applied to bands below the dropout band. However, in our simulated data analysis, we find that the requirement of SNR(F090W)$<2$ rejects $\sim75\%$ of $z\sim2$ galaxies that would otherwise have leaked in as interlopers. In addition, the SNR(F090W) criterion helps reject $\sim70\%$ of $6<z<7$ sources that would have otherwise been included in the sample. Including the F090W SNR criterion has minimal effect on the candidates with true redshift $z>7.25$, and merely shifts the average redshift of the Z dropouts from 7.2 to 7.5. 

Based on these \texttt{JAGUAR} tests, the color and SNR selections yield a sample with purity $\gtrsim95\%$ for objects brighter than F444W$\leq27.5$ mag. Here we define purity as a ratio of the number of candidates with true redshift $z>7$ to the number of all selected candidates. The purity drops quickly to $\sim50\%$ at F444W=28.5 mag, where the contamination by $z\sim2$ galaxies is about 30\%. The other 20\% of the objects that passed the selection criteria are $5<z<7$ galaxies. Finally, we test the selection criteria on the mock NIRCam images of the observed field described in \citetalias{Merlin2022}. We find a similar conclusion in terms of both purity and completeness. 
\begin{figure}
    \centering
    \includegraphics[width=0.45\textwidth]{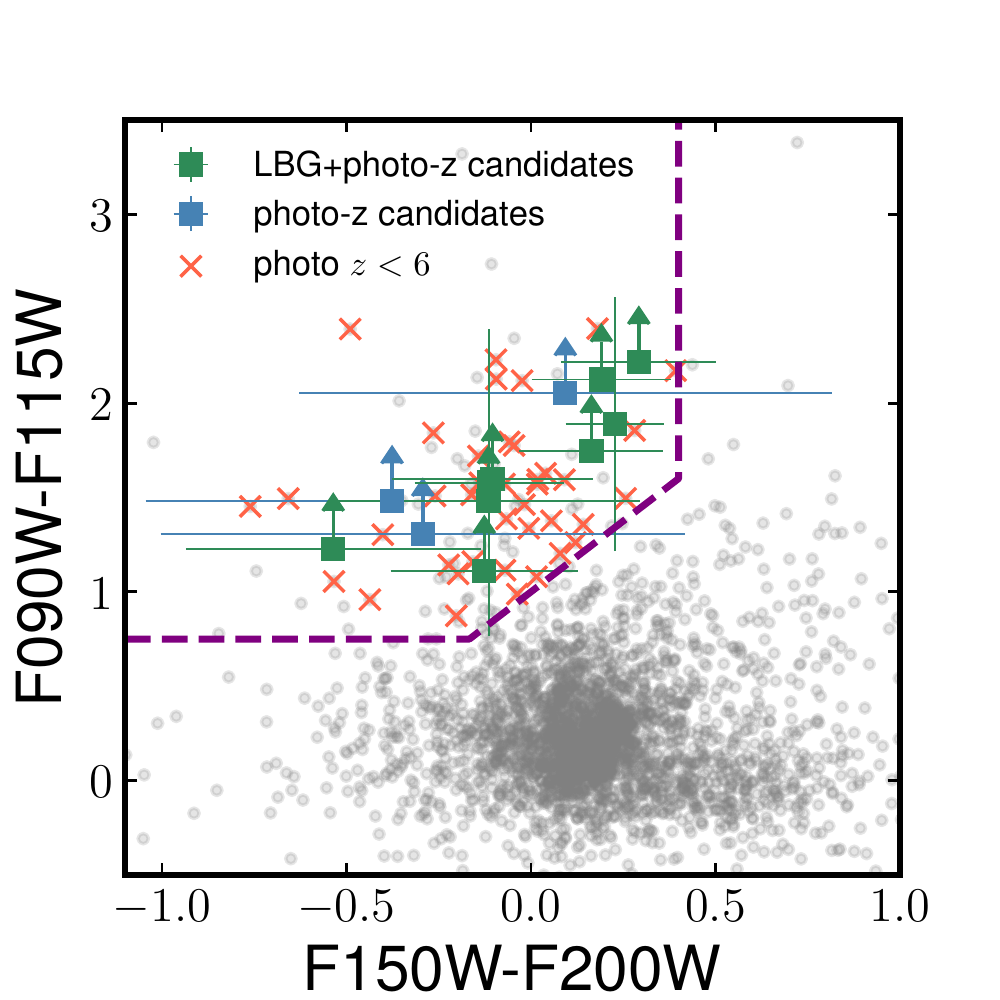}
    \caption{Observed color selection diagram for F090W dropouts. Green squares indicate the \emph{LBG$+$photo-z} candidates in this paper  (see Section \ref{sec:selection_criteria}). Orange crosses are objects that pass the color and SNR selection criteria but failed the photo-z criterion. The gray dots show all objects in the photometric catalog that pass the minimum SNR criteria (i.e. SNR(F444W)$>$10 and SNR(F115W, F150W, F200W)$>2$). The gray dots that are in the color selection box are those with SNR(F090W)$>2$ and do not pass the SNR(F090W) criterion. Blue squares are \emph{photo-z} candidates (see Section \ref{sec:photoz_candidates}).}
    \label{fig:color-diagram}
\end{figure}

\subsection{Photometric redshift refinement}
To further refine the sample selection, we follow \citet{Kauffmann2020}, who found that adding a photometric redshift (photo-z) selection to a color cut yields higher purity with a minimal loss of completeness. Thus, we refine the color-SNR selected sample with the photometric redshift code \texttt{EAzY} \citep{Brammer2008} with the default V1.3 spectral template and a flat prior. 

Based on the output photometric redshifts, we restrict the selection to candidates (1) with the most probable redshift at high $z$ and (2) with an integrated probability under the peaks of the distribution that is largest at high $z$, i.e. $z_{peak}>6$ and $z_p>6$ in \texttt{EAzY} nomenclature. In the test on the \texttt{JAGUAR} mock catalog, this procedure reduces the interloper contamination from $\sim30\%$ to $\lesssim10\%$ at F444W$=28.5$ mag. The completeness drops by an additional few percent for sources with F444W brighter than 27 mag, by $\sim 20\%$ at 28 mag, and by $\sim 40\%$ at 28.5 mag.

As for candidates from the actual JWST images, we have a total of 55 candidates from the color and SNR cut alone. We plot the color selection diagram of these galaxies in Figure \ref{fig:color-diagram}. The photo-$z$ refinement rejects 41 objects. Among those that fail the photo-$z$ criterion, 25 are either artefacts, partially fall into gaps, or blobs with visible sizes larger than $1\arcsec$. For the rest, most have fluxes in the F090W band with SNR$>1$ and are fainter than F444W$>27$ mag, except for three           sources which could well be at $z>7$ based on colors and visual inspection but are still excluded from further analysis for consistency of the selection criteria. As a result, we have 14 candidates for further refinement. We call this type of candidates \emph{LBG$+$photo-z} sample. They are our primary sample.
 
\subsection{Additional photometric candidates}\label{sec:photoz_candidates}
In addition to the \emph{LBG$+$photo-z} sample above, we perform a second independent selection to capture additional high-$z$ candidates that fail to enter into the strict LBG selection window. This additional sample may include the sources within the color-color region co-populated by high redshift sources and passive intermediate redshift interlopers, and the sources that may be faint in the F115W to F200W bands. To do so, we carry out a photometric redshift analysis of all objects in the source catalog and select galaxies that have best-fit photometric redshifts in the range of $7<z<9$ assessed by two photometric redshift codes: \texttt{EAzY} and \texttt{zphot} \citep{Fontana2000}. We refer readers to \citet[Paper XI]{Santini2022} for further detail of the \texttt{zphot} fitting procedure. Furthermore, we require that their $0.1\arcsec$ central regions must not fall in gaps in at least 6 out of the 7 bands, including both F090W and F115W, which we deem critical as those bands encompass the Lyman break. Lastly, at least three of the bands from F115W to F356W must have SNR$>2$, and F444W must have SNR$\gtrsim8$. Based on mock-catalog analysis, having photometric redshift confirmed by two independent codes increases the purity of the sample. Based on simulated NIRCam images of the observed field, we estimate a sample purity of $>70\%$ for this additional photometric selection. 

We find 14 candidates, nine of which are already included in the \emph{LBG$+$photo-z} candidates.  
Therefore, we have 5 remaining objects for further refinement. We identify these objects as the \emph{photo-z} sample. 

\subsection{Contamination by cool dwarfs}
Although the expected number of ultracool dwarfs is only a few per field \citep{Ryan_Jr2016}, we take a precautionary step to remove possible contamination from late L and T dwarfs in our sample. One effective solution to mitigate this contamination, without resorting to characterization of the light profile of the sources to determine whether they are spatially resolved or point sources, is to exclude objects with F356W$-$F444W$<-0.3$ \citep{Hainline2020}. With this criterion, we reject 2 additional objects from the \emph{LBG$+$photo-z} sample and none from the \emph{photo-z} sample.

\subsection{Visual Inspection}
Finally, two authors (NL and MT) visually reviewed all candidates. The objective of this process is strictly limited to eliminating defects and artefacts that could escape automatic flagging, such as spikes of foreground stars, scattered light and objects that fall into gaps. Such a sanity check is especially important  for a new camera and telescope. To be conservative, we only include objects that both reviewers independently approved. Here we remove 3 objects from the \emph{LBG$+$photo-z} sample. One is an artefact known as `Dragon's breath' in the JWST artefact nomenclature \citep{Rigby2022}. One partially falls in a gap in F090W, and one is a large faint blob next to bright objects. We remove one object from the \emph{photo-z} sample that partially falls in the gap in the F090W image.

\begin{figure*}
\centering
\includegraphics[width=0.195\textwidth]{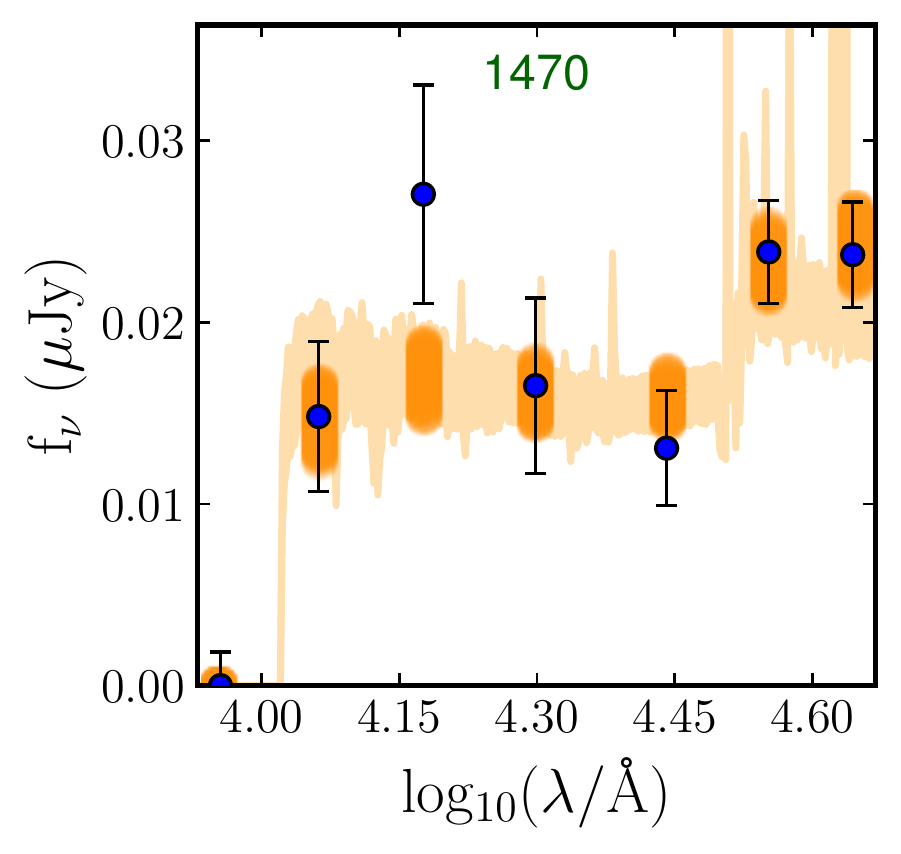}
\includegraphics[width=0.195\textwidth]{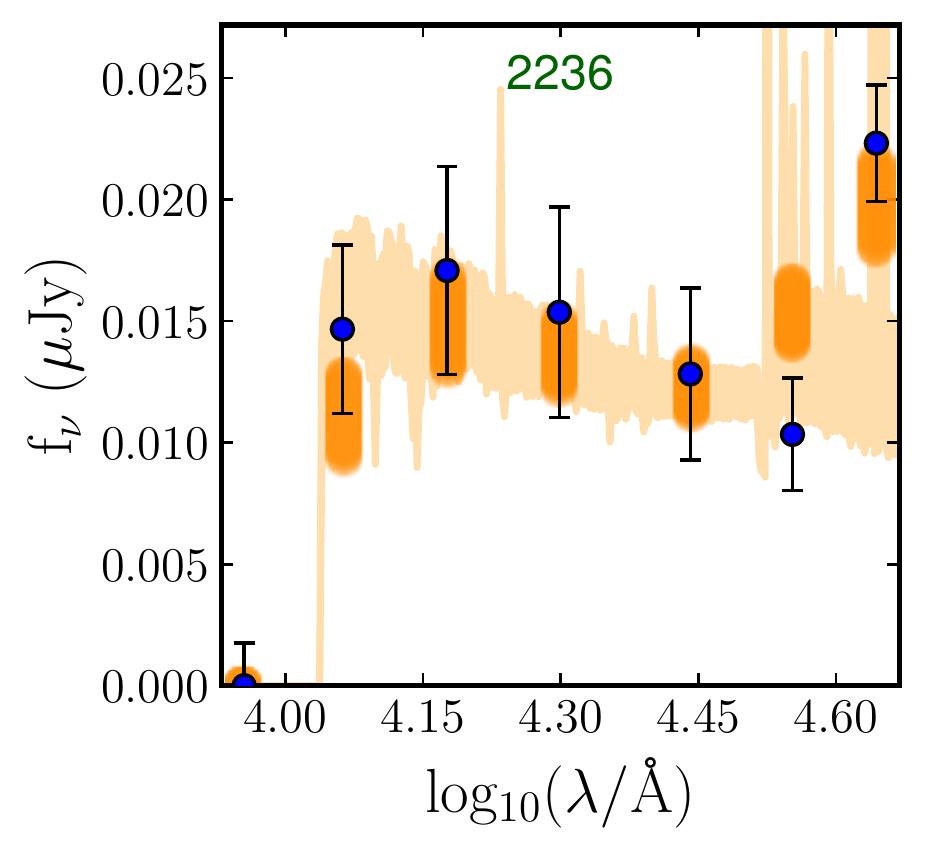}
\includegraphics[width=0.195\textwidth]{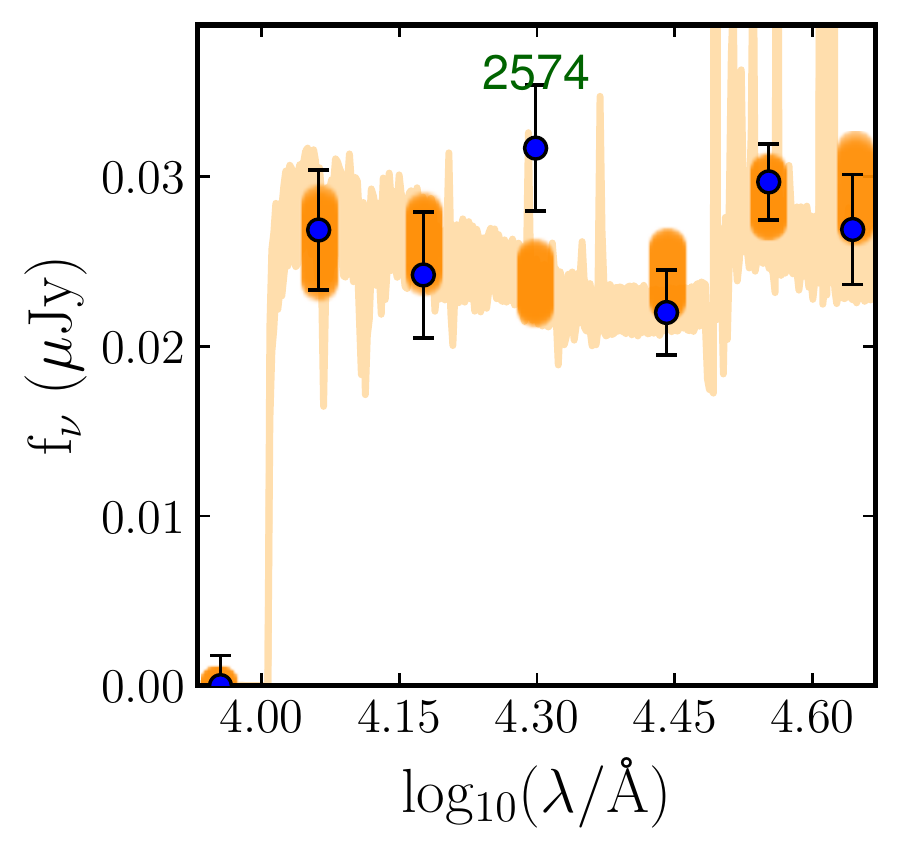}
\includegraphics[width=0.195\textwidth]{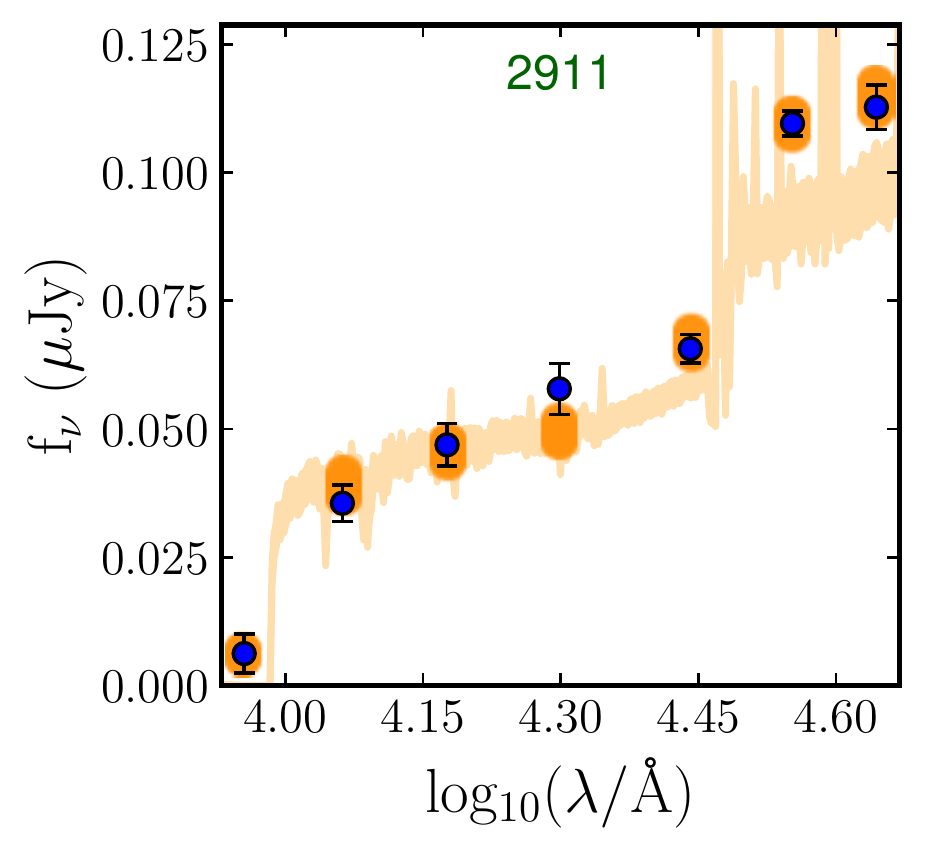}
\includegraphics[width=0.195\textwidth]{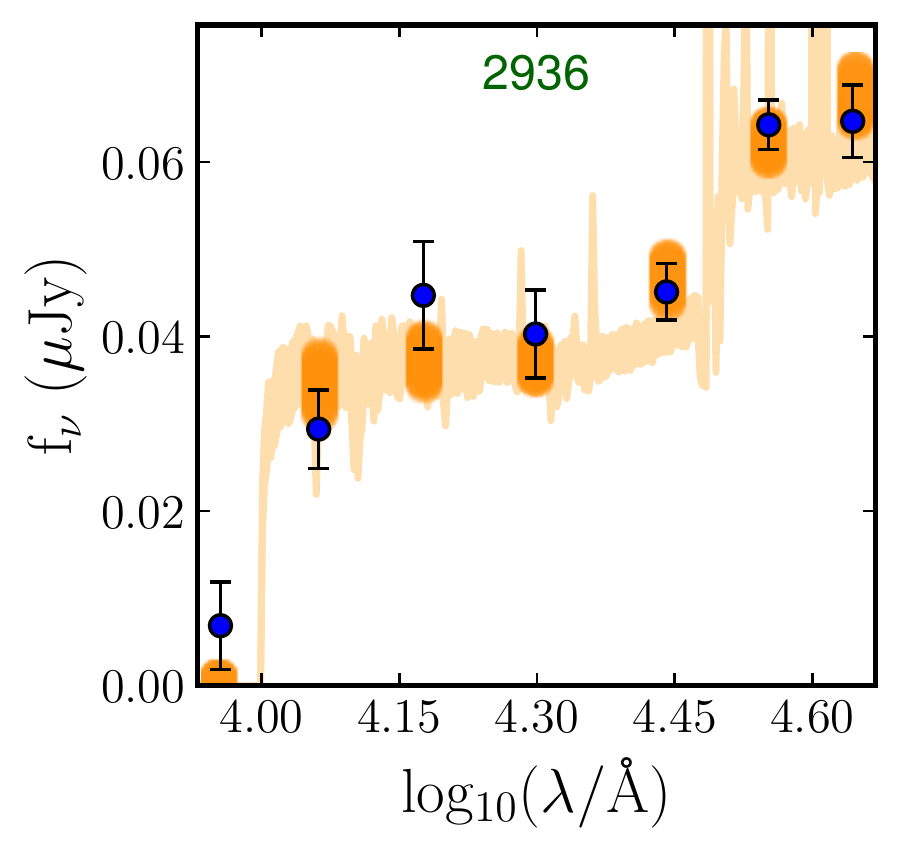}\\
\includegraphics[width=0.195\textwidth]{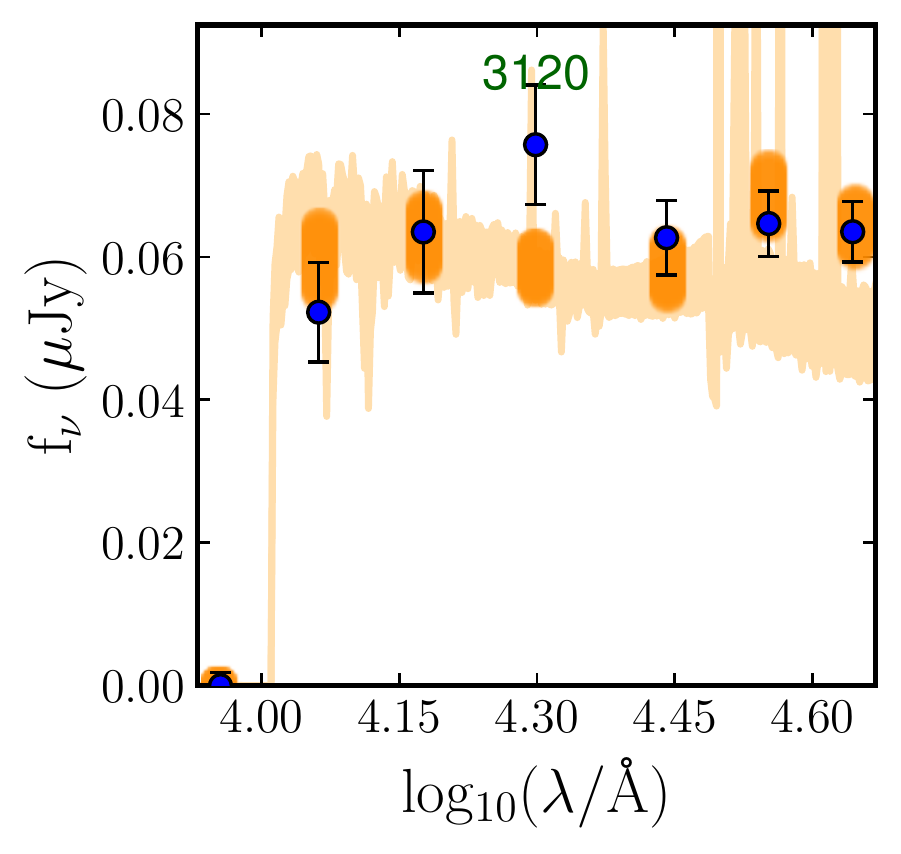}
\includegraphics[width=0.195\textwidth]{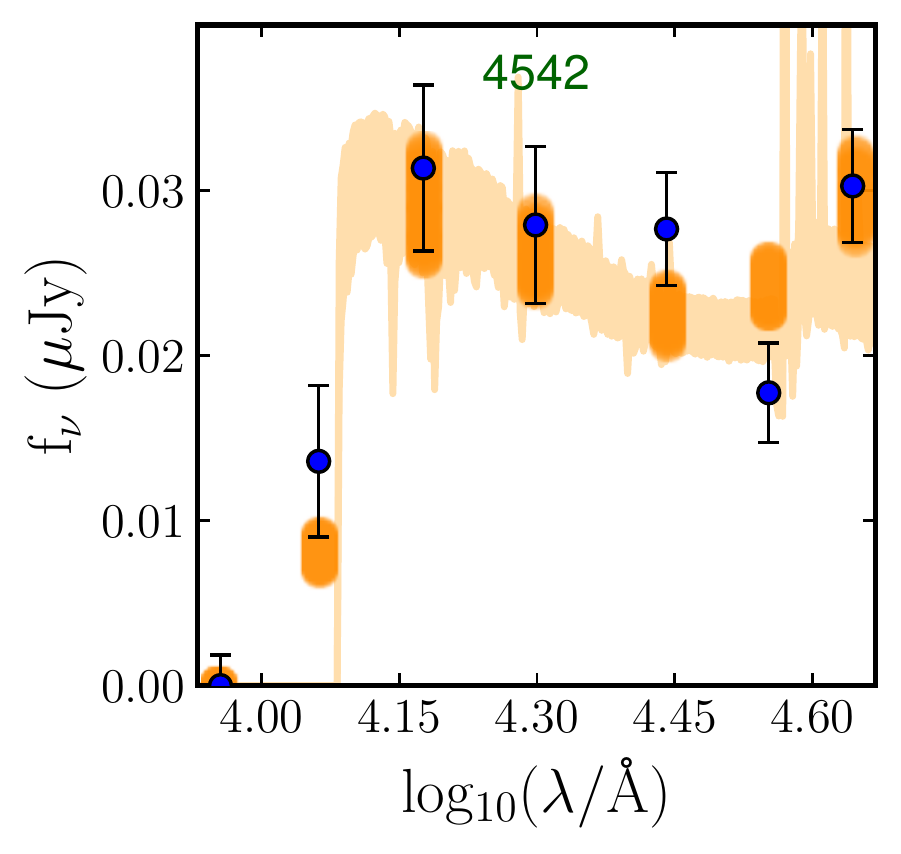}
\includegraphics[width=0.195\textwidth]{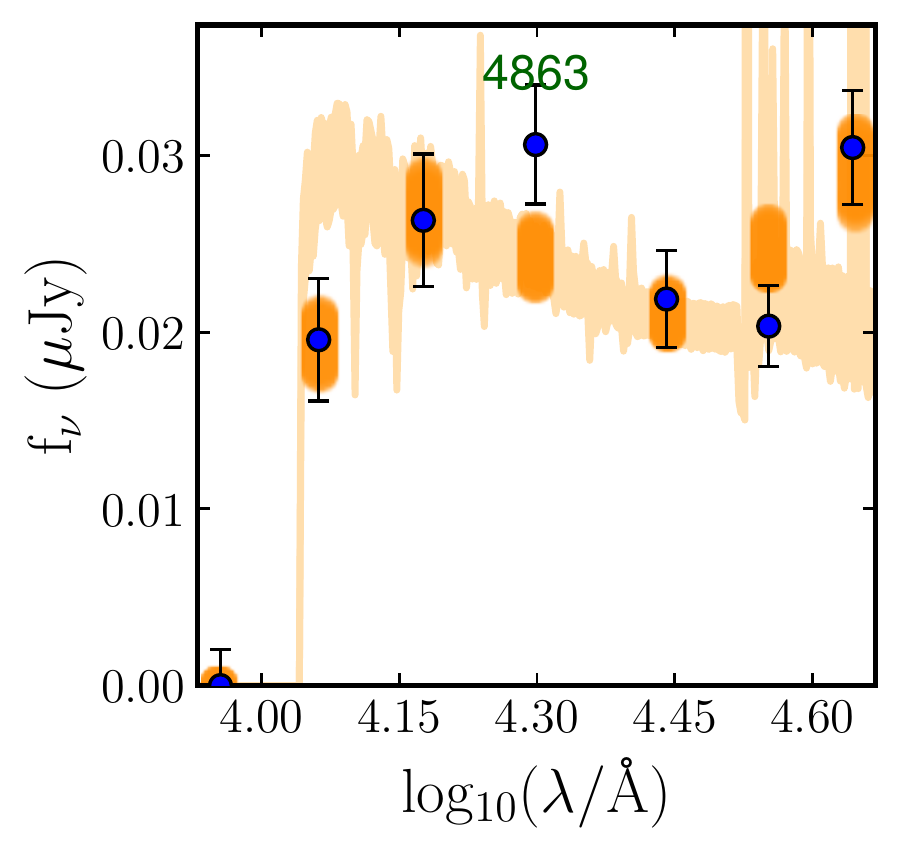}
\includegraphics[width=0.195\textwidth]{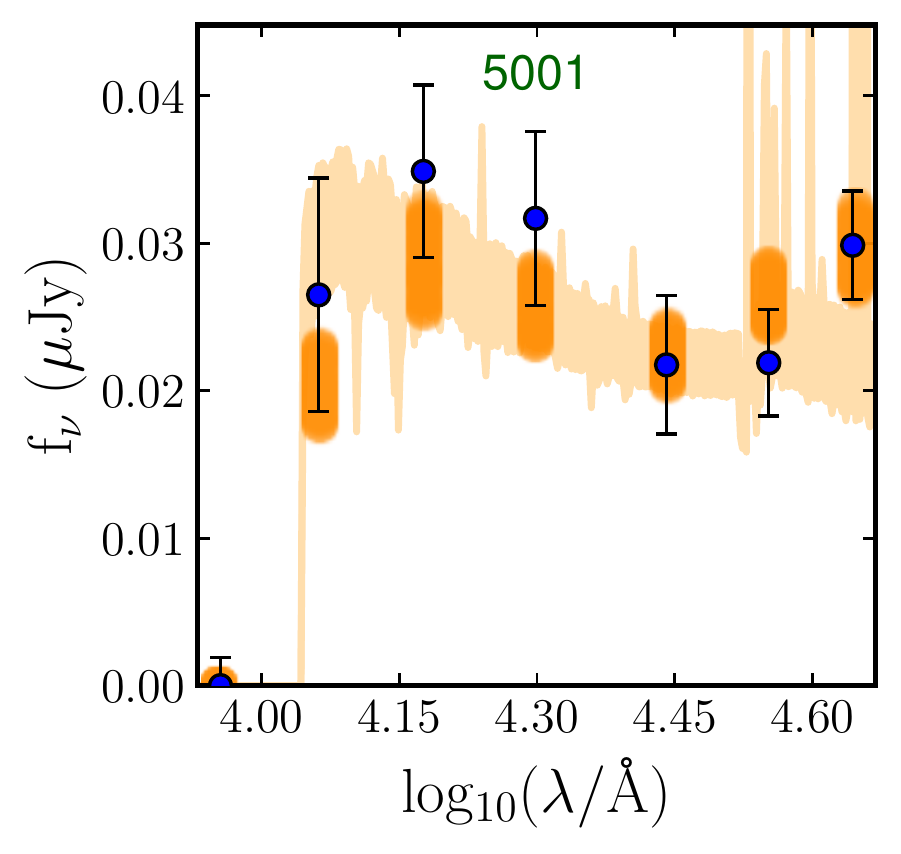}
\includegraphics[width=0.195\textwidth]{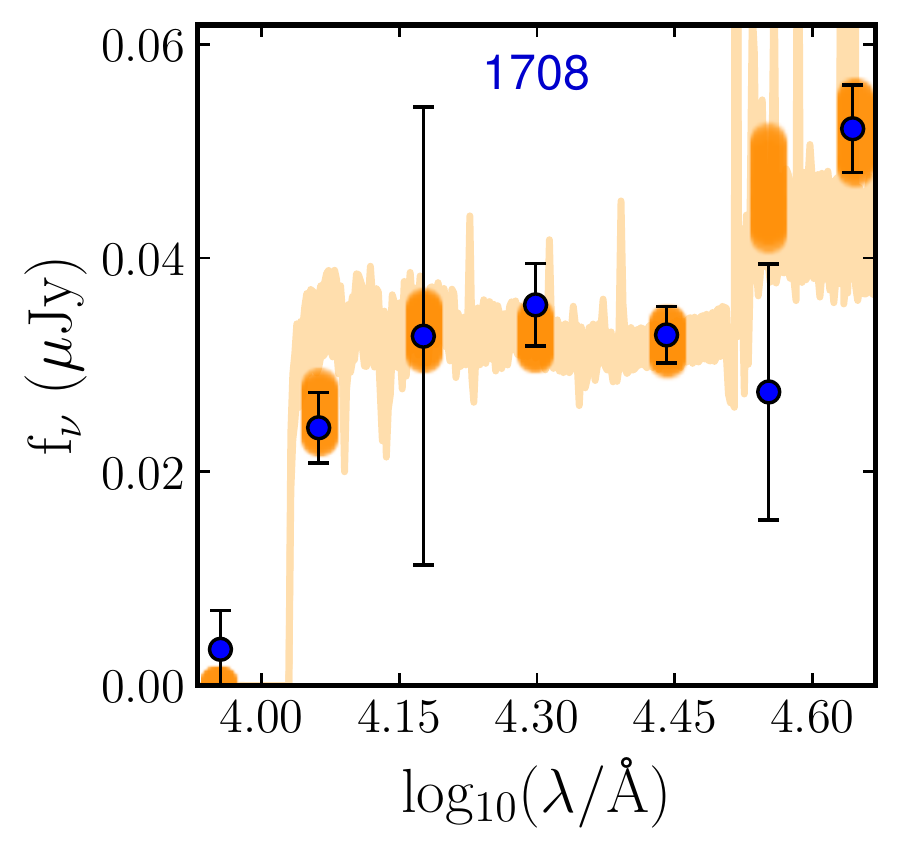}\\
\includegraphics[width=0.195\textwidth]{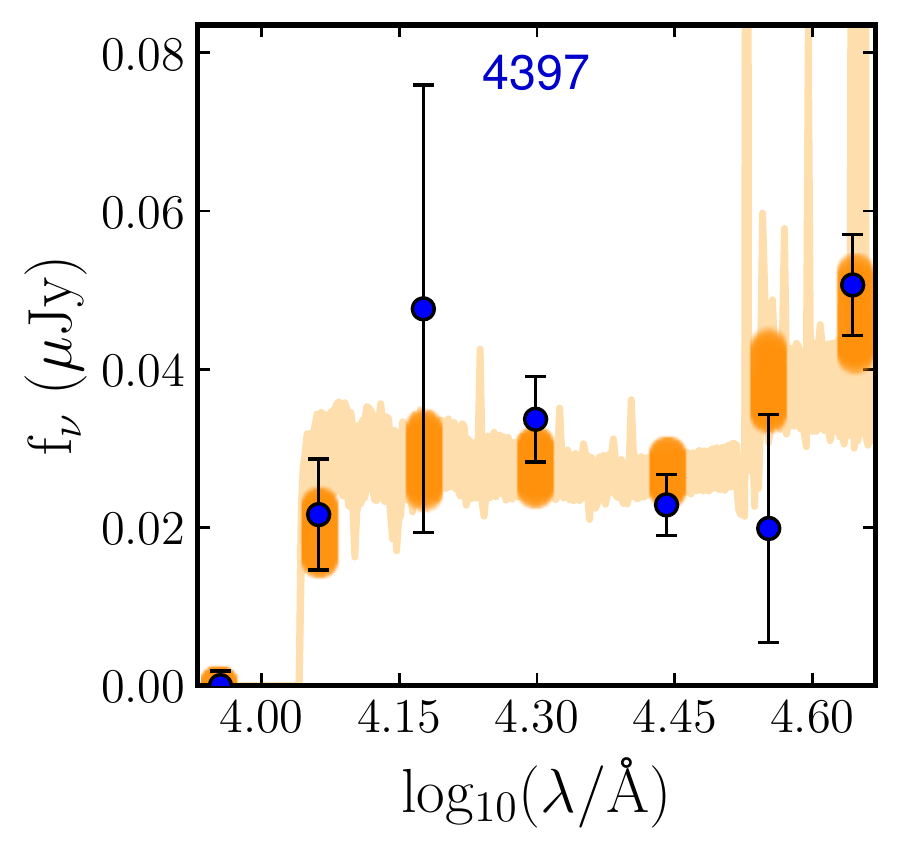}
\includegraphics[width=0.195\textwidth]{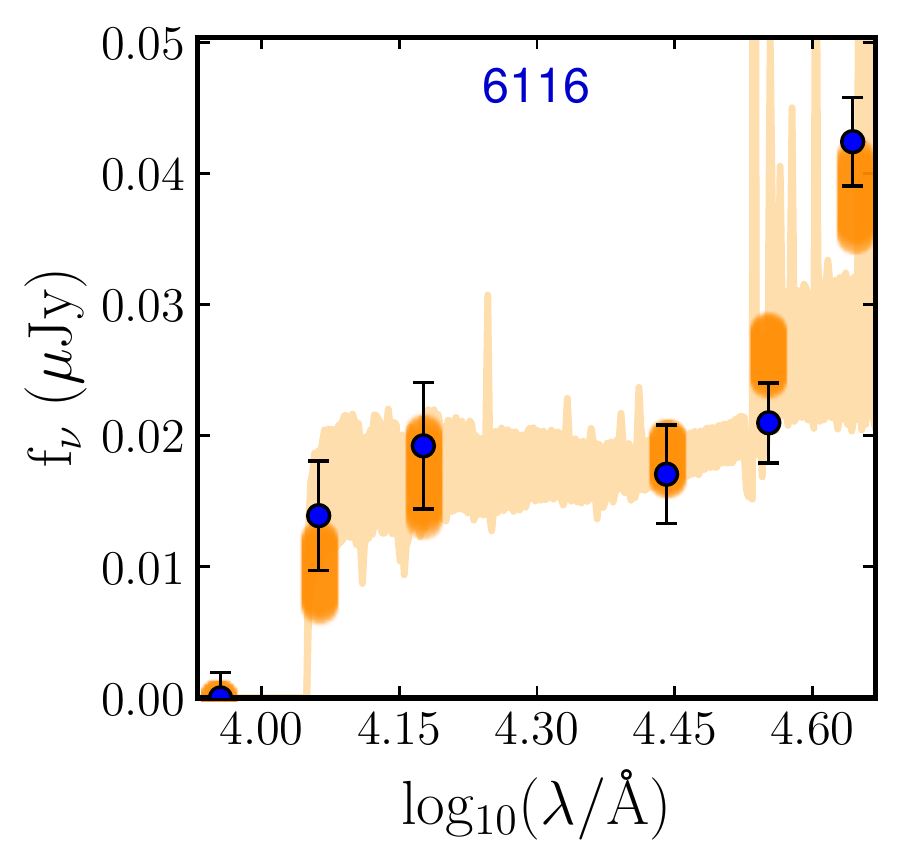}
\includegraphics[width=0.195\textwidth]{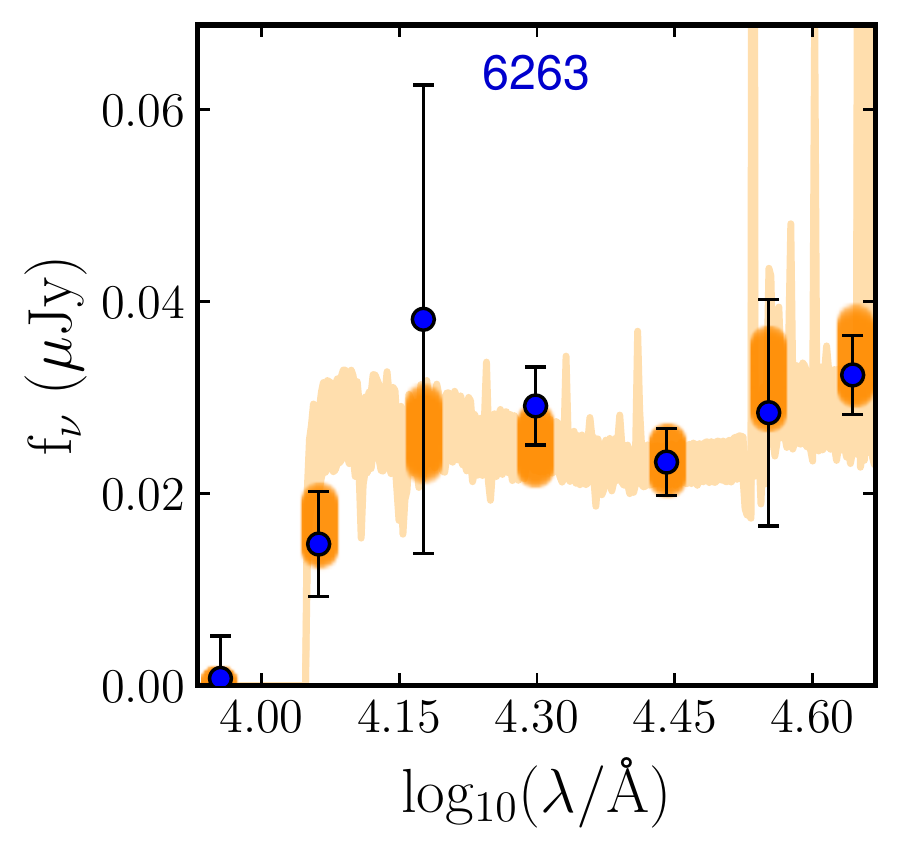}
\caption{Best-fit SEDs of $7\lesssim z \lesssim 9$ candidates, derived with the most probable redshift ($z_{peak}$) determined by \texttt{EAzY}. Observed total fluxes are shown in blue. The first 9 objects with IDs in green are the \emph{LBG$+$photo-z} candidates. The last 4 objects with IDs in blue are the \emph{photo-z} candidates. \label{fig:SEDs}}
\end{figure*}

\section{Results and discussion}
We obtain 9 final \emph{LBG$+$photo-z} candidates and 4 final \emph{photo-z} candidates at $8\sigma$-detection in F444W. Their image stamps are shown Figure \ref{fig:image_stamps}. We discuss their photometric properties in Appendix \ref{ap:photometric}. The F444W magnitude range of the candidates are 26.3 to 28.0 with a median of 27.6 mag. We discuss their SED-inferred physical properties in the sections below. Other properties such as mass-to-light ratios, morphology and sizes are discussed in detail by \citetalias{Santini2022}, \citet[Paper XII]{Treu2022b}, and \citetalias{Yang2022} respectively. 

The total area where the bands required for candidates selections (F090W, F115W, F150W, F200W and F444W) overlap is 7.7 arcmin$^2$. Based on the segmentation map, 87\% are not covered by foreground objects. Our preliminary test on mock catalogs suggest that the completeness of the \emph{LBG$+$photo-z} selection is $\sim40\%$. Using the existing UV luminosity functions \citep{Bouwens2015,Bowler2020}, we expect $\sim10$ \emph{LBG$+$photo-z} candidates at $7<z<9$ in the field, which is in agreement with the number of candidates found here. Since previous UV luminosity functions were mainly based on rest-frame UV-selected galaxies, our results suggest that the population of LBG galaxies that are selected from the rest-frame V band are similar to those selected from the rest-frame UV band.  Nonetheless, this comparison is only qualitative because we do not have detailed source recovery and completeness simulations available for this preliminary version of the image reduction.

\subsection{SED Modeling}
We use the Bayesian SED-fitting code \texttt{Bagpipes} \citep{Carnall2018} to infer physical properties of the galaxies in our sample. The code is based on the 2016 version of the \citet{BC03} stellar population models, the \texttt{CLOUDY} photoionization code \citep{Ferland2017}, the \citet{Inoue2014} IGM attenuation model, and the \citet{Kroupa2002} IMF. We assume the \citet{Calzetti2000} dust attenuation law. The free parameters and their prior ranges that are not related to SFH are stellar mass [$6<\log(M_*/M_\odot)<13$], attenuation in the V band [$0<A_V<3$], ionization parameter [$-4<\log U<-2$], and metallicity $[0<Z/Z_\odot<1]$. 

While parameters derived from SED fitting are relatively independent of model templates \citep{Whitler2022},
fitting codes and assumed dust attenuation law \citep{Topping2022}, they are dependent on the assumed SFH \citep{Carnall2019, Leja2019, Topping2022,Whitler2022}. The dependence is stronger for the age and weaker for the stellar mass and SFR \citep{Santini2015,Tacchella2022}. For this reason, we adopt a log-normal SFH, which allows a wide range of SFH shapes ranging from rising to declining over time, despite its parametric nature. Moreover, this functional form matches well the overall shapes of the majority of SFHs in cosmological hydrodynamic simulations \citep[e.g.,][]{Gladders2013,Diemer2017}. The free parameters under this assumption are the age of the universe at the peak of star formation $t_\textrm{peak}$ and the FWHM of the SFH duration. We set both to vary between 1 Myr to 13 Gyr. 

For each object, we run \texttt{Bagpipes} twice. First, with a redshift fixed to the $z_{peak}$ redshift from \texttt{EAZY}. The best-fit values of the inferred physical properties quoted in this paper are from this step. The redshift is fixed to the \texttt{EAZY} best-fit values because 1) \texttt{EAZY} is a common photometric redshift SED fitting tool that has been used and validated in multiple deep surveys \citep[e.g.,][]{Straatman16,Skelton2014} and 2) for consistency with other papers in the series that use \texttt{EAZY} output as the reference photometric redshifts e.g., \citet[Paper XV]{Glazebrook2022} and \citet[Paper XVI]{Nanayakkara2022}. For the second run of \texttt{Bagpipes}, we allow the redshift to vary within $3\sigma$ of $z_{peak}$. The uncertainties reported in this paper account for redshift uncertainties from this step. The best-fit SEDs at $z_{peak}$ are shown in Figure \ref{fig:SEDs}. We list the inferred physical properties in Table \ref{tab:properties}.

The stellar mass of our candidates ranges from $10^{8.0}$ to $10^{9.3} M_\odot$ with a median of $10^{8.5} M_\odot$. The mass-weighted age ranges from 35 Myr to 220 Myr with a median of 140 Myr.
The ages correspond to a formation redshift (the redshift at which half of the current stars in each galaxy have formed) of $z=10.6$ to $z=7.8$, with a median of $z=9.5$.
This suggests that the majority of these galaxies started forming stars before redshift 9.5.
For the more massive galaxies with ages close to 200 Myr, we infer substantial star formation happening at $z\gtrsim11$, which is consistent with the $z>11$ objects found in the same dataset (\citetalias{Castellano2022},\ \citealp{Naidu2022}). The inferred dust extinction ($A_V$) has an average of $0.2^{+0.2}_{-0.1}$ mag, which is consistent with $A_V\sim0.5$ mag of the $z=8.38$ A2744\_YD4 lensed galaxy \citep{Laporte2017}, as modelled by \citet{Behrens2018}. This average extinction is a few times lower than typical values at $z\sim2-3$ \citep[e.g.,][]{Theios2019}.

Our best-fitted SFHs all show SFRs that increase with time, in approximate agreement with what is found in cosmological simulations at similar mass and redshift without burstiness features \citep[e.g.,][]{Ma2018}. As a cautionary note, based on \citet{Whitler2022}, about $\lesssim10\%$ of the $z\sim7$ galaxies are undergoing an intense burst of star formation. For these galaxies, the assumed smooth SFH would be skewed toward the recent burst, likely resulting in an underestimation of mass and age. 

\begin{deluxetable*}{lcccccccc}
\tablecaption{The physical properties inferred from the F090W dropout galaxies \label{tab:properties}}
\tablecolumns{9}
\tablewidth{0pt}
\tablehead{
\colhead{ID} & \colhead{RA} & \colhead{DEC} &
\colhead{z(\texttt{EAzY})} & \colhead{$\log(M_*/M_\odot)$} & \colhead{SFR ($M_\odot$/yr)} & \colhead{Age (Myr)} &\colhead{$M_{UV}$\tablenotemark{a}} & \colhead{$A_V$}}

\startdata
\multicolumn{9}{c}{\emph{LBG$+$photo-z} candidates }\\
1470 & 3.5137326 & -30.3628201 & $7.6^{+0.5}_{-0.5}$ & $8.4^{+0.2}_{-0.1}$ & $1.4^{+0.4}_{-0.2}$ & $160^{+76}_{-55}$ & $-18.7^{+0.2}_{-0.3}$ & $0.2^{+0.2}_{-0.1}$ \\
2236 & 3.4898570 & -30.3544532 & $8.0^{+0.6}_{-0.7}$ & $8.0^{+0.3}_{-0.7}$ & $0.9^{+0.3}_{-0.6}$ & $72^{+88}_{-67}$ & $-18.7^{+0.3}_{-0.1}$ & $0.1^{+0.2}_{-0.1}$ \\
2574 & 3.4954699 & -30.3507722 & $7.4^{+0.3}_{-0.3}$ & $8.4^{+0.1}_{-0.3}$ & $1.6^{+0.3}_{-0.3}$ & $140^{+48}_{-95}$ & $-19.2^{+0.1}_{-0.2}$ & $0.1^{+0.1}_{-0.1}$ \\
2911 & 3.5118088 & -30.3468413 & $6.9^{+0.2}_{-0.2}$ & $9.3^{+0.2}_{-0.1}$ & $9.1^{+3.6}_{-1.9}$ & $180^{+81}_{-69}$ & $-19.5^{+0.1}_{-0.2}$ & $0.7^{+0.1}_{-0.1}$ \\
2936 & 3.5119322 & -30.3466618 & $7.2^{+0.6}_{-5.0}$ & $9.1^{+0.1}_{-0.2}$ & $4.2^{+1.5}_{-0.9}$ & $230^{+74}_{-120}$ & $-19.4^{+0.1}_{-0.5}$ & $0.4^{+0.2}_{-0.2}$ \\
3120 & 3.5202581 & -30.3439199 & $7.4^{+0.3}_{-0.3}$ & $8.5^{+0.3}_{-0.2}$ & $3.2^{+1.3}_{-1.3}$ & $35^{+71}_{-22}$ & $-20.1^{+0.1}_{-0.3}$ & $0.2^{+0.1}_{-0.2}$ \\
4542 & 3.4879853 & -30.3254241 & $9.0^{+0.7}_{-0.6}$ & $8.5^{+0.2}_{-0.3}$ & $2.0^{+0.4}_{-0.5}$ & $110^{+42}_{-63}$ & $-19.5^{+0.2}_{-0.1}$ & $0.1^{+0.1}_{-0.1}$ \\
4863 & 3.4866961 & -30.3272162 & $8.1^{+0.4}_{-0.5}$ & $8.1^{+0.3}_{-0.3}$ & $1.4^{+0.4}_{-0.7}$ & $52^{+64}_{-35}$ & $-19.3^{+0.1}_{-0.1}$ & $0.1^{+0.1}_{-0.0}$ \\
5001 & 3.4997064 & -30.3177257 & $8.1^{+0.7}_{-0.8}$ & $8.3^{+0.2}_{-0.4}$ & $1.6^{+0.4}_{-0.7}$ & $74^{+67}_{-52}$ & $-19.4^{+0.2}_{-0.1}$ & $0.1^{+0.1}_{-0.0}$ \\
\multicolumn{9}{c}{\emph{photo-z} candidates }\\
1708 & 3.4905553 & -30.3603869 & $7.8^{+0.4}_{-0.6}$ & $8.8^{+0.2}_{-0.2}$ & $3.3^{+0.8}_{-0.7}$ & $150^{+68}_{-79}$ & $-19.5^{+0.1}_{-0.3}$ & $0.3^{+0.1}_{-0.2}$ \\
4397 & 3.4746591 & -30.3226226 & $8.1^{+0.6}_{-0.6}$ & $8.8^{+0.2}_{-0.2}$ & $2.6^{+1.3}_{-0.5}$ & $160^{+74}_{-82}$ & $-19.4^{+0.3}_{-0.2}$ & $0.2^{+0.3}_{-0.1}$ \\
6116 & 3.5045910 & -30.3079217 & $8.2^{+0.6}_{-0.6}$ & $8.7^{+0.2}_{-0.7}$ & $2.6^{+1.4}_{-1.4}$ & $140^{+96}_{-120}$ & $-18.8^{+0.4}_{-0.2}$ & $0.5^{+0.3}_{-0.2}$ \\
6263 & 3.4696869 & -30.3090295 & $8.2^{+0.5}_{-5.7}$ & $8.5^{+0.2}_{-0.4}$ & $2.1^{+0.7}_{-1.9}$ & $130^{+630}_{-74}$ & $-19.3^{+3.6}_{-0.2}$ & $0.1^{+0.3}_{-0.1}$\\
\enddata
\tablenotetext{a}{$M_\textrm{UV}$ magnitudes are measured by sampling best-fit SEDs from the posterior distributions of the best-fit parameters. They are magnitudes corresponding to the average flux in the $100$\angstrom\ range centering at 1500\angstrom\ wavelength. }
\tablecomments{We model the observed photometry with \texttt{BAGPIPES} assuming a log-normal SFH. The values are median of the posterior distribution when redshift is fixed to $z_{peak}$ from \texttt{EAZY}. The uncertainties include the marginalized $1\sigma$ interval when redshift is free to vary (see text). The age displayed is mass-weighted age. All galaxies are assumed to have magnification $\mu=1$.} 
\end{deluxetable*}

\subsection{SFR of $7<z<9$ galaxies}
\begin{figure*}
    \centering
    \includegraphics[width=0.95\textwidth]{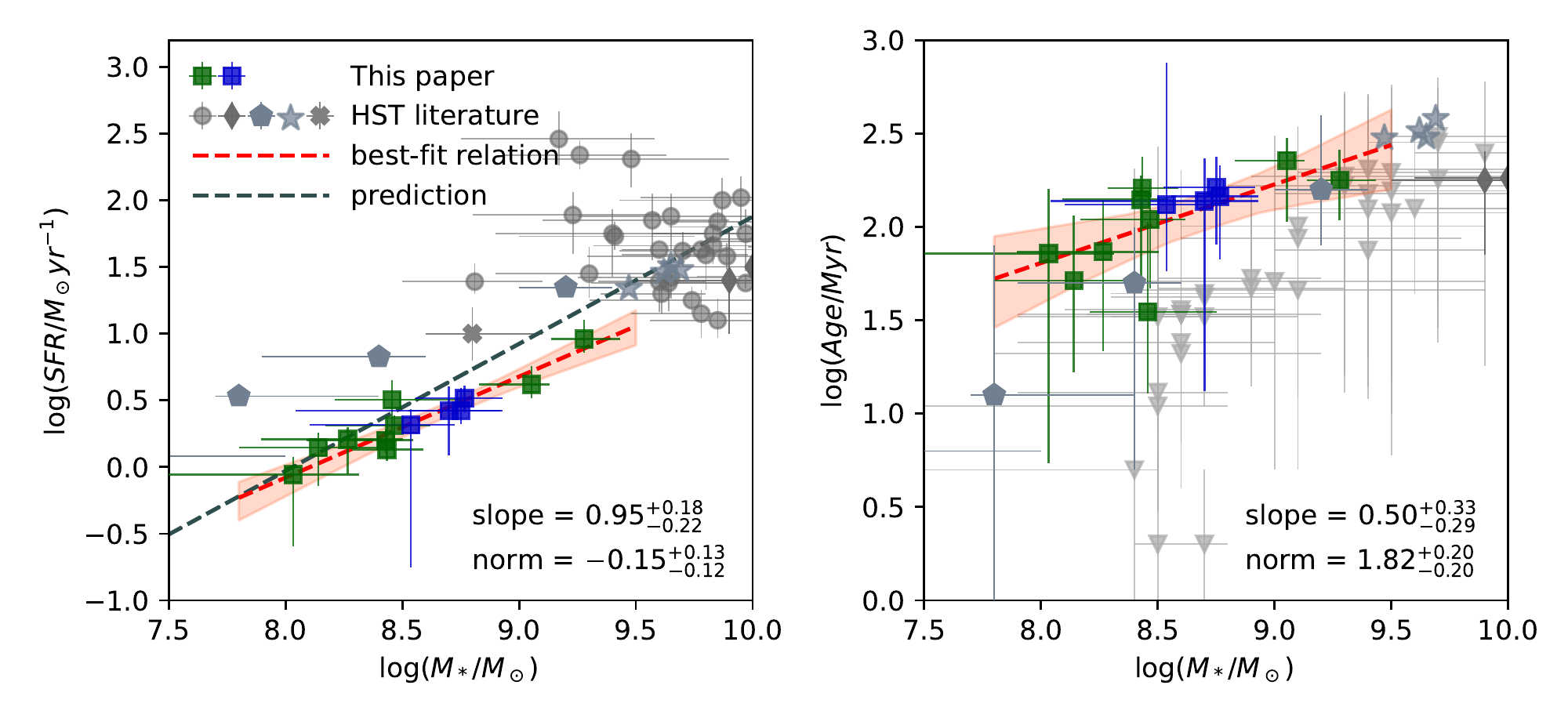}
    \caption{Left: star-formation main sequence (log SFR vs log stellar mass). Right: log mass-weighted age vs log stellar mass. Green and blue squares show our \emph{LBG$+$photo-z} and \emph{photo-z} candidates in this paper, respectively. The best-fit observed linear relation is plotted in red in each panel, while the Eddington-bias corrected slope and normalization at $10^8M_\odot$ are listed at the lower right corners. The gray data points show literature values. Circles -- rest-frame UV selected $z\sim7$ galaxy with ALMA FIR data (assumed non-parametric SFHs) from \citet{Topping2022}. Diamond -- $z\sim8.5$ galaxies in CANDELS fields with Spitzer/IRAC data (assumed non-parametric SFHs) from \citet{Tacchella2022}. Pentagons -- stacked $z\sim8$ galaxies in the HST legacy fields with deep Spitzer/IRAC data (assumed constant SFHs) from \citet{Stefanon2022a_stack}. Stars and filled cross -- $z\sim8$ galaxies with IRAC detections from SuperBoRG (assumed delayed-tau and non-parameteric SFHs respectively) \citep{Roberts-Borsani2022,Morishita2020}. Triangles -- $z\sim6.8$ galaxies in the COSMOS field with Spitzer/IRAC data (assumed constant SFHs) from \citet{Whitler2022}. Gray dash line indicates theoretical prediction at $z=8$ from \citet{Yung2019}.}
    \label{fig:mass_age_SFR}
\end{figure*}

We plot the star-formation main sequence in the left panel of Figure \ref{fig:mass_age_SFR}. Our measurements are shown in colors. Gray data points show the measurements in literature based on SED modeling with nebular emission lines.
We fit our measurements with a linear relation in log scale using a least squares method and Monte Carlo sampling (repeated 1000 times):
\begin{equation}
    \log_{10} (\frac{\textrm{SFR}}{M_{\odot}~\mathrm{yr^{-1}}}) = \alpha\log_{10}(\frac{M_*}{10^8 M_\odot})+\beta,
\end{equation}
where $\alpha$ and $\beta$ are free parameters. The best-fit slope and normalization at $10^{8}M_\odot$ are $0.76^{+0.15}_{-0.14}$ and $-0.08^{+0.10}_{-0.14}$, with an average scatter of 0.1 dex.

We follow the method in \citet{Santini2017} to correct the fitted linear equation for Eddington Bias. In short, we replicate the observed mass 20 times to increase the statistics and consider these as `true' masses. We then create a grid of true $\alpha$'s and $\beta$'s. For each `true' $\alpha$ and $\beta$, we calculate each galaxy's `true' main-sequence SFR with the observed scatter. We then add Gaussian noise based on each object's measurement uncertainties to obtain the `observed' mass and SFR. We calculate the `observed' best-fit linear fit for that `true' $\alpha$ and $\beta$. The Eddington-bias corrected parameters are the range of these true $\alpha$'s and $\beta$'s that match the observed slope and normalization and their uncertainty ranges. We find that the Eddington-bias corrected parameters are $\alpha= 0.95^{+0.17}_{-0.23}$ and $\beta = -0.15^{+0.13}_{-0.12}$.

The obtained main-sequence slope $\alpha$ is consistent with lower redshifts \citep{Santini2017}. Both hydrodynamical simulations and semi-analytical models predict $\alpha$ that is close to unity, e.g. $\alpha=1.03$ in \citet{Ma2018}, and $\alpha=0.95$ in \citet{Yung2019}. Both are consistent with our measurement.

\subsection{SFR evolution insight from galaxy ages}
Age as a function of stellar mass provides additional information to constrain the relation between star formation and mass at even higher redshifts, where it might be difficult to obtain large enough samples from direct observations, even with JWST. Here we present a method to constrain the time-dependent component in the main-sequence evolution. We assume that SFR ($\dot M$) is a power-law function of both stellar mass and redshift:
\begin{equation}
    \dot M \propto M^{\alpha} (1+z)^{\gamma}  \propto M^{\alpha} (t)^{\gamma\prime}
\label{eq:sfr}
\end{equation}
$M$ and $t$ are stellar mass and time since the Big Bang. In the matter-dominated universe (which is relevant to the calculation here), $\gamma\prime=-\frac{2}{3}\gamma$. Since mass-weighted age can be approximated as
\begin{equation}
    \textrm{age}(t_0) \approx \frac{\int_0^{t_0}\dot M (t_0-t)dt}{\int_0^{t_0}\dot Mdt},
\end{equation}
where $t_0$ is the age of the universe at the observed redshift, we can use equation \ref{eq:sfr} to show that
\begin{equation}
    \textrm{age}(M) \propto M^{\frac{1-\alpha}{1+\gamma\prime}}.
\end{equation}
Here we predict a linear relation between log mass-weighted age and log stellar mass. With this formalism, we can constrain the time-dependent term of the star formation rate $\gamma\prime$ from its slope and the $\alpha$ term obtained in the above section. We note that this approach is approximate under the assumption that the galaxies have not started their quenching process and have been, on average, residing on the main sequence, i.e., no significant burst that produced a substantial fraction of stellar mass.

We plot $\log$ age vs $\log$ stellar mass in the right panel of Figure \ref{fig:mass_age_SFR}. We obtained a slope of $0.42^{+0.25}_{-0.25}$, a normalization of $1.81^{+0.18}_{-0.21}$, and a scatter of 0.2 dex. Using the same method to correct for the Eddington bias, we infer an intrinsic slope and normalization of $0.50^{+0.33}_{-0.29}$ and $1.82^{+0.21}_{-0.20}$. This indicates that the redshift-dependent term for the SFR evolution function is $\gamma=1.4^{+0.7}_{-0.8}$. Our result is roughly consistent within $1\sigma$ with the predicted value of $\gamma=2.01$ based on the FIRE-II simulation \citep{Ma2018}.

We note that ages measured on star-forming galaxies should always be treated as estimates only. Age is sensitive to multiple parameters such as the assumed SFH, metalliticity, and binary fractions \citep[e.g.,][]{Steidel2016,Leethochawalit2019}. This is particularly true in galaxies with high specific star formation rate \citep{Whitler2022}. Such galaxies may include those with red F444W$-$F356W color, e.g. ID4863 and ID6116 in our sample, which are potentially due to high-EW nebular emission lines \citep[][]{Roberts-Borsani2016}, indicating highly active star-forming activities.

\section{Conclusion}
We present 13 sources at $7<z<9$, photometrically selected from the NIRCam observation in the GLASS-JWST ERS program. The galaxies have been identified based on rest-frame V-band detection, the first time that this has been possible for a sample of $L\lesssim L_*$ sources. Based on simulated data modeling we expect our sample to have $\lesssim10\%$ contamination. We measured their stellar mass, SFRs, and mass-weighted ages using the \texttt{Bagpipes} SED fitting code. The galaxies have stellar masses ranging from $10^{8.0}$ to $10^{9.3}M_\odot$. The key results from this work are: 
\begin{itemize}
    \item Overall, the number density of these sources is broadly consistent with expectations from the UV luminosity function at this redshift determined from HST data. However, a detailed comparison is beyond the scope of this work as it requires exhaustive completeness and source recovery simulations. 
 
    \item The galaxies span a wide range in mass-weighted ages, from 30 Myr in the least massive galaxy to 100-200 Myr in the more massive $10^{9}M_\odot$ galaxies. This suggests significant amount of star formation in typical objects in our sample at redshift $z\gtrsim 11$. A caveat to note is that these conclusions are based on an assumed log-normal functional form for the SFH. 
    
    \item We present a preliminary determination of the star-formation main sequence at $z\sim 8$. The Eddington-bias corrected slope is $0.95^{+0.17}_{-0.23}$, consistent within $1 \sigma$ with predictions from \citet{Ma2018} and \citet{Yung2019}.
    \item We present a method to constrain the star-formation main sequence evolution using inferred ages. We find that the redshift-dependent power-law slope term is $1.4^{+0.7}_{-0.8}$.
\end{itemize}

We note that the galaxies in this work may be affected by systematic uncertainties due to the preliminary NIRCam calibrations available at the time of writing. The fluxes were obtained at face values from the catalog in \citetalias{Merlin2022}, where the uncertainties in the zero-points ($\sim0.1$ mag) are not included in the photometry calculations. Also, the field is affected by some modest amount of gravitational lensing, which would impact both mass and SFR measurements, but not age. The effect should be relatively minor for the star-formation main sequence since our inference is presented in a log scale. We will consider the lensing magnification in future work after the full data acquisition of the GLASS-ERS program.

Overall, this paper provides a first robust look of the properties of typical galaxies in the epoch of reionization from 19 hours of observations. It highlights the transformational capabilities of JWST infrared imaging to characterize the first light sources in the universe once both deeper and wider datasets become available.

\section*{Acknowledgments}
This work is based on observations made with the NASA/ESA/CSA James Webb Space Telescope. The data were obtained from the Mikulski Archive for Space Telescopes at the Space Telescope Science Institute, which is operated by the Association of Universities for Research in Astronomy, Inc., under NASA contract NAS 5-03127 for JWST. These observations are associated with program JWST-ERS-1324. We acknowledge financial support from NASA through grants JWST-ERS-1324.  This research is supported in part by the Australian Research Council Centre of Excellence for All Sky Astrophysics in 3 Dimensions (ASTRO 3D), through project number CE170100013. KG and TN acknowledge support from Australian Research Council Laureate Fellowship FL180100060. MB acknowledges support from the Slovenian national research agency ARRS through grant N1-0238.

\section*{Data Availability}
We provide tables in this paper in machine readable format. Further images, plots, and the routines used in this work are available upon reasonable request. 

\vspace{5mm}
\facilities{JWST(NIRCam)}
\software{astropy \citep{astropy:2018},  
          Bagpipes \citep{Carnall2018}}



\appendix
\restartappendixnumbering
\section{Observed photometric properties of the Z dropouts}\label{ap:photometric} 
We list photometric properties of our candidates in Table \ref{tab:photometry}. Their F444W magnitudes range from 26.3 to 28.0 with an average of $27.4^{+0.4}_{-0.5}$, while the F150W magnitudes range from 27.2 to 28.0 with an average of $27.6\pm0.4$ mag. At $7<z<8.5$, the F277W band is generally free from strong emission lines and can be used to probe the continuum below the 4000\angstrom\ break. However, $H\beta+$[OIII] lines fall in the observed F444W band while the 4000\angstrom\ break generally falls in the 356W band. There is no emission line-free band above the 4000\angstrom\ break. The average F277W$-$F444W of our sample is $0.5^{+0.2}_{-0.3}$ mag, suggesting either a presence of the 4000\angstrom\ break (and thus a presence of intermediate-age $>0.3$ Gyr stars) or strong $H\beta +$[OIII] emission lines.

\begin{deluxetable*}{lrrrrrrrrrr}[b!]
\tablecaption{Summary of photometry of F090W dropout galaxies\label{tab:photometry}}
\tablecolumns{11}
\tablewidth{0pt}
\tablehead{
\colhead{ID} &
\colhead{F090W} & \colhead{F115W} & \colhead{F150W} & \colhead{F200W} &
\colhead{F277W} & \colhead{F356W} & \colhead{F444W} & \colhead{$m_\textrm{F444W}$} &
\colhead{color1\tablenotemark{a}} &  \colhead{color2\tablenotemark{b}}}
\startdata
\multicolumn{11}{c}{\emph{LBG$+$photo-z} candidates }\\
1470 & $0.0\pm1.8$ & $14.8\pm4.1$ & $27.0\pm6.0$ & $16.5\pm4.8$ & $13.1\pm3.2$ & $23.8\pm2.8$ & $23.7\pm2.9$  & 28.0 & -0.5 & $>1.2$ \\
2236 & $0.0\pm1.8$ & $14.6\pm3.5$ & $17.1\pm4.3$ & $15.3\pm4.3$ & $12.8\pm3.5$ & $10.3\pm2.3$ & $22.3\pm2.4$  & 28.0 & -0.1 & $>1.5$ \\
2574 & $0.0\pm1.8$ & $26.8\pm3.5$ & $24.2\pm3.7$ & $31.6\pm3.7$ & $22.0\pm2.5$ & $29.6\pm2.2$ & $26.8\pm3.2$  & 27.8 & 0.3 & $>2.2$ \\
2911 & $6.2\pm3.8$ & $35.5\pm3.5$ & $46.9\pm4.1$ & $57.8\pm4.9$ & $65.6\pm2.8$ & $109.4\pm2.4$ & $112.6\pm4.3$  & 26.3 & 0.2 & 1.9 \\
2936 & $6.9\pm5.0$ & $29.4\pm4.5$ & $44.7\pm6.1$ & $40.2\pm5.0$ & $45.0\pm3.2$ & $64.2\pm2.8$ & $64.6\pm4.1$  & 26.9 & -0.1 & 1.6 \\
3120 & $0.0\pm1.9$ & $52.2\pm7.0$ & $63.4\pm8.6$ & $75.6\pm8.3$ & $62.6\pm5.2$ & $64.6\pm4.6$ & $63.5\pm4.2$  & 26.9 & 0.2 & $>2.1$ \\
4542 & $0.0\pm1.8$ & $13.6\pm4.6$ & $31.3\pm5.0$ & $27.9\pm4.7$ & $27.6\pm3.4$ & $17.7\pm3.0$ & $30.2\pm3.4$  & 27.7 & -0.1 & $>1.1$ \\
4863 & $0.0\pm2.0$ & $19.5\pm3.5$ & $26.3\pm3.7$ & $30.6\pm3.4$ & $21.8\pm2.7$ & $20.3\pm2.3$ & $30.4\pm3.2$  & 27.7 & 0.2 & $>1.7$ \\
5001 & $0.0\pm1.9$ & $26.5\pm7.9$ & $34.8\pm5.8$ & $31.6\pm5.9$ & $21.7\pm4.7$ & $21.9\pm3.6$ & $29.8\pm3.7$  & 27.7 & -0.1 & $>1.6$ \\
\multicolumn{11}{c}{\emph{photo-z} candidates }\\
1708 & $3.4\pm3.6$ & $24.1\pm3.3$ & $32.6\pm21.4$ & $35.6\pm3.9$ & $32.8\pm2.6$ & $27.4\pm12.0$ & $52.0\pm4.1$  & 27.1 & 0.1 & $>2.1$ \\
4397 & $0.0\pm1.9$ & $21.6\pm7.0$ & $47.6\pm28.3$ & $33.6\pm5.4$ & $22.8\pm3.8$ & $19.8\pm14.4$ & $50.6\pm6.4$  & 27.1 & -0.4 & $>1.5$ \\
6116 & $0.0\pm1.9$ & $13.9\pm4.2$ & $19.2\pm4.8$ & nan & $17.0\pm3.8$ & $21.0\pm3.0$ & $42.4\pm3.4$  & 27.3 & 45.4 & $>1.1$ \\
6263 & $0.8\pm4.4$ & $14.7\pm5.5$ & $38.1\pm24.4$ & $29.1\pm4.0$ & $23.3\pm3.5$ & $28.4\pm11.8$ & $32.3\pm4.1$  & 27.6 & -0.3 & $>1.3$ \\
\enddata
\tablenotetext{a}{F150W$-$F200W UV continuum color}
\tablenotetext{b}{F090W$-$F115W Lyman drop color}
\tablecomments{The fluxes are total fluxes in nJys. When the total flux is negative, we set the flux to zero in the SED fitting procedure and set the uncertainty to the flux of $1\sigma$ limiting magnitude. Colors are based on aperture magnitudes. For color calculation, the F090W aperture magnitudes are set to 1$\sigma$ limiting magnitudes if the SNR is less than one and thus the F090W-F115W colors appear as lower limits. The IDs match those in \citetalias{Merlin2022}.}
\end{deluxetable*}

\begin{figure*}
\setlength{\lineskip}{-6pt}
\begin{center}
\includegraphics[width=0.8\textwidth]{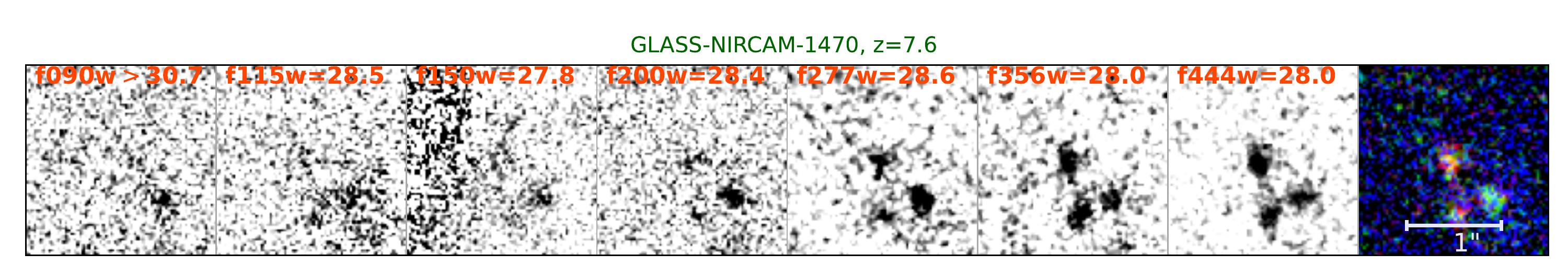}\\
\includegraphics[width=0.8\textwidth]{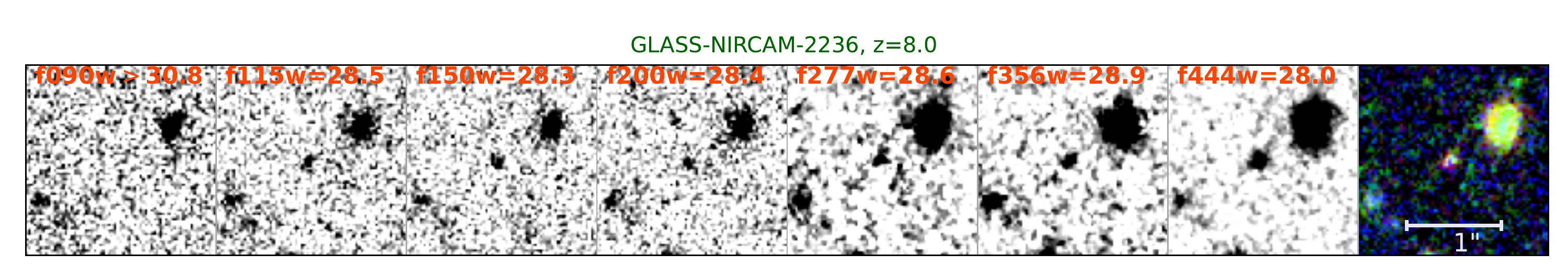}\\
\includegraphics[width=0.8\textwidth]{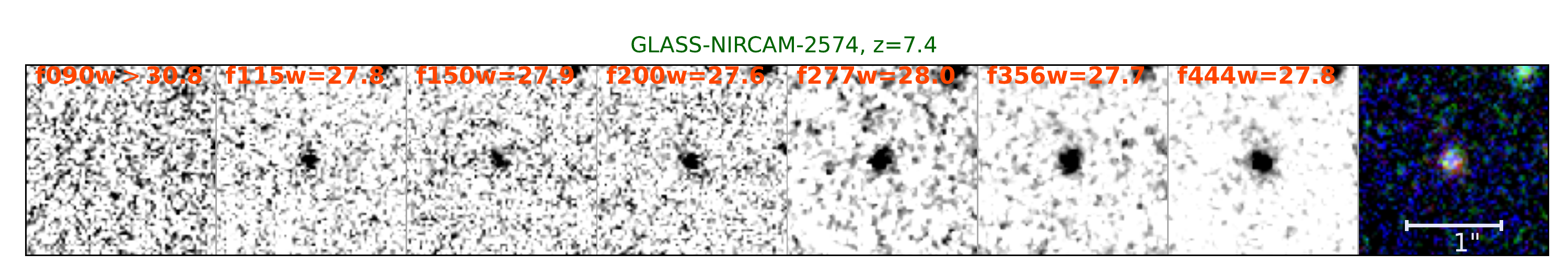}\\
\includegraphics[width=0.8\textwidth]{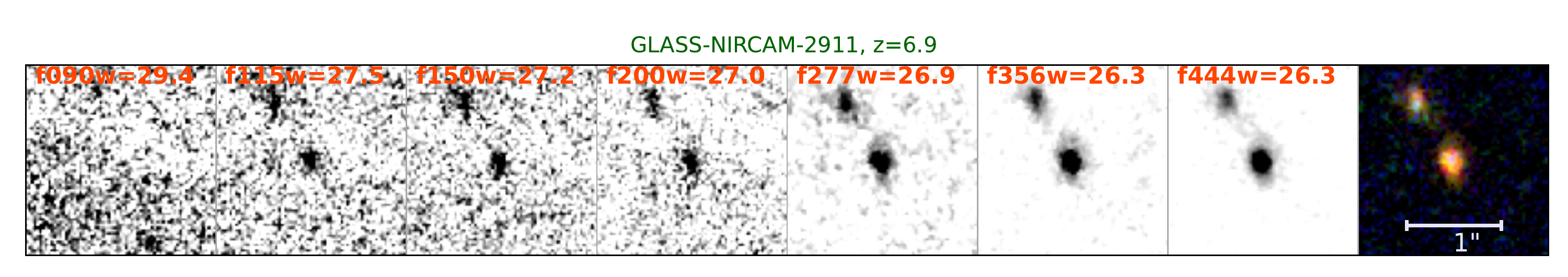}\\
\includegraphics[width=0.8\textwidth]{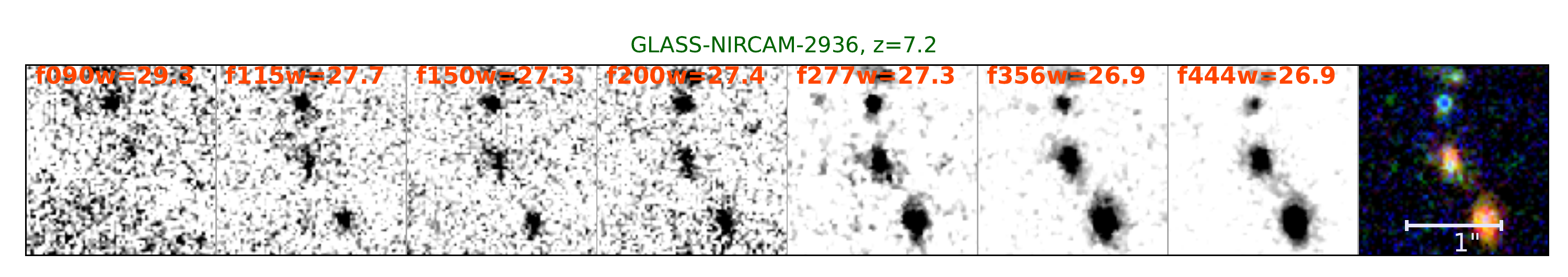}\\
\includegraphics[width=0.8\textwidth]{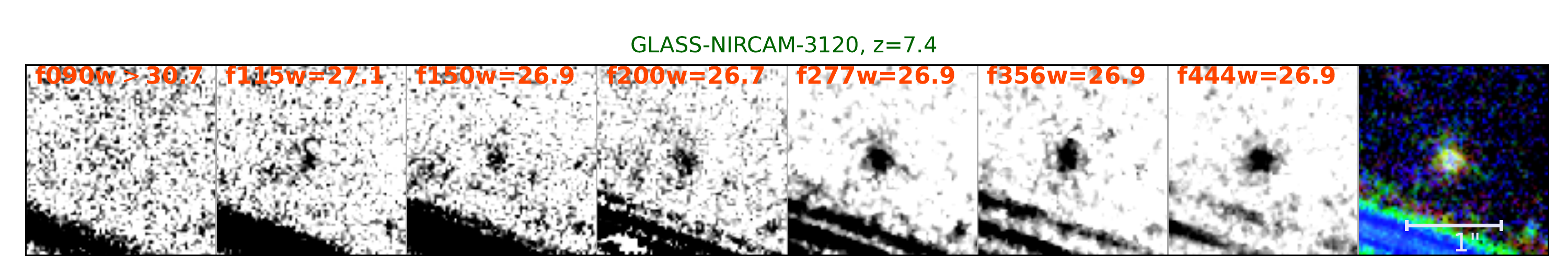}\\
\includegraphics[width=0.8\textwidth]{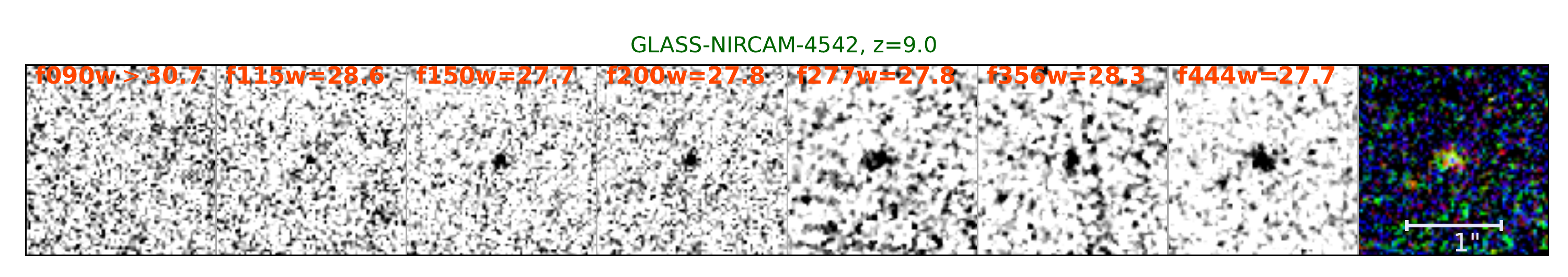}\\
\includegraphics[width=0.8\textwidth]{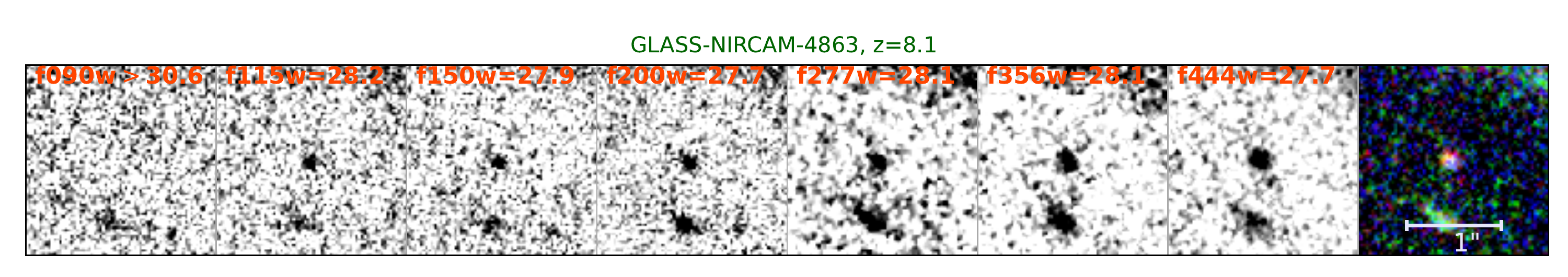}\\
\includegraphics[width=0.8\textwidth]{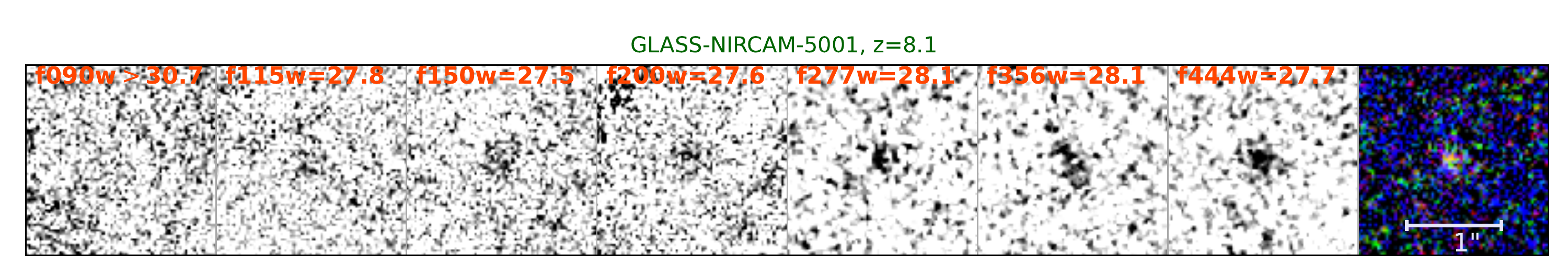}
\end{center}
\caption{Image stamps of $7<z<9$ \emph{LBG$+$photo-z} candidates found in the GLASS-JWST ERS NIRCam parallel observation at SNR(F444W)$>10$. The columns from left to right are the images in F090W, F115W, F150W, F200W, F277W, F356W, F444W, and RGB composite based on the F444W, F277W, and F115W band.  \label{fig:image_stamps}}
\end{figure*}

\begin{figure*}
\setlength{\lineskip}{-6pt}
\begin{center}
\includegraphics[width=0.8\textwidth]{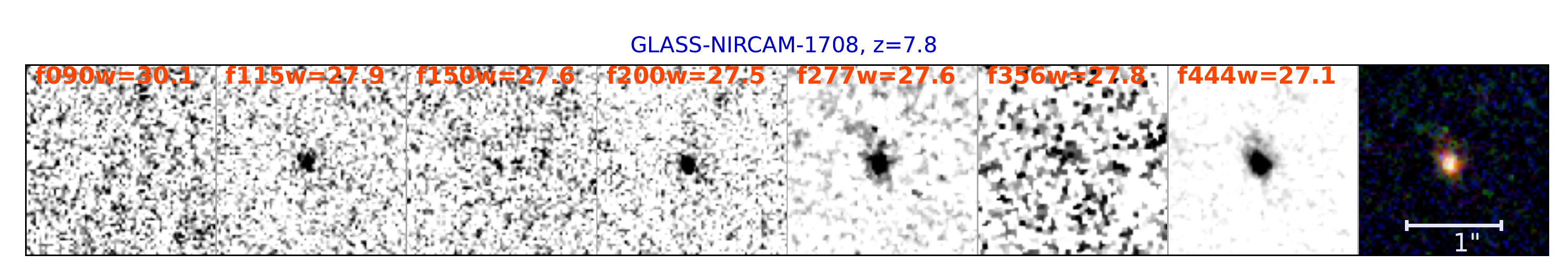}\\
\includegraphics[width=0.8\textwidth]{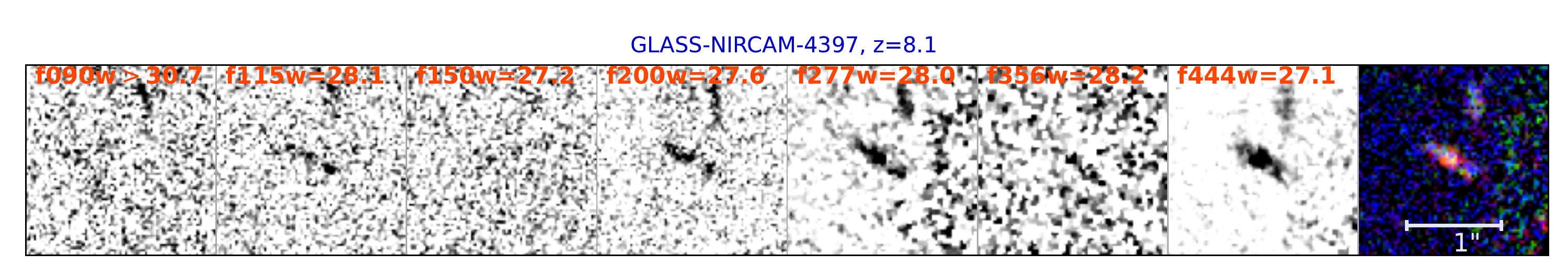}\\
\includegraphics[width=0.8\textwidth]{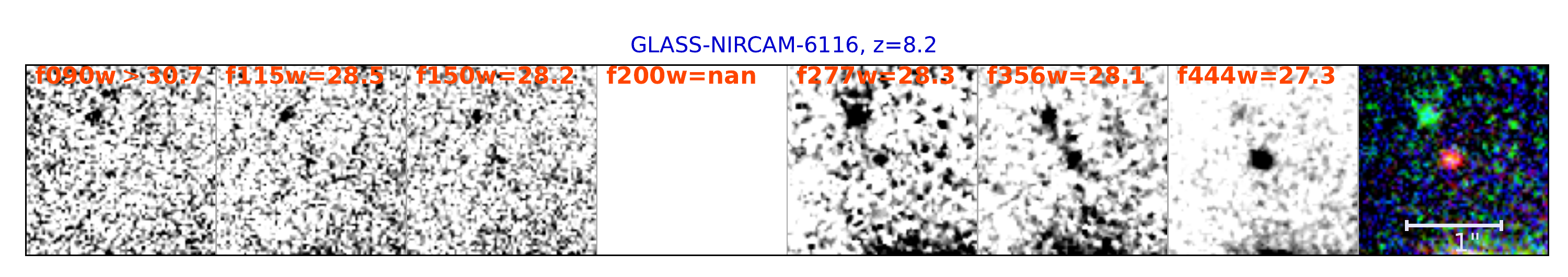}\\
\includegraphics[width=0.8\textwidth]{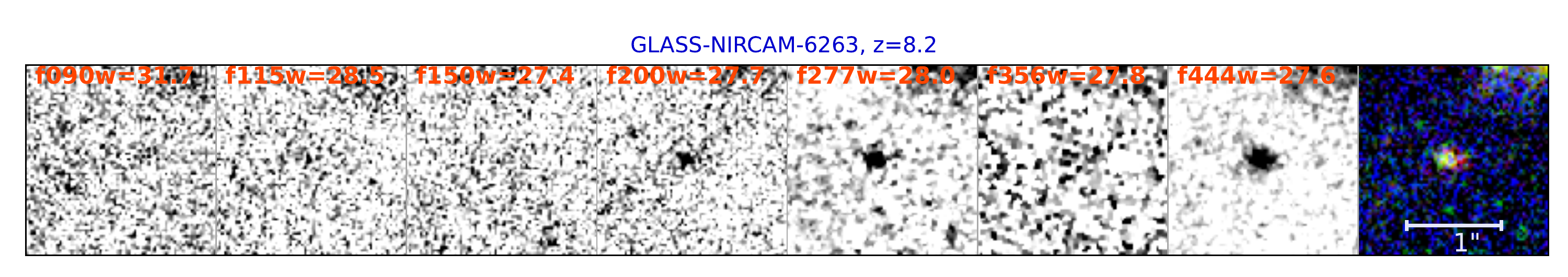}
\end{center}
\caption{Same as Figure \ref{fig:image_stamps} but for the \emph{photo-z} candidates.}
\end{figure*}

\bibliography{ref}{}

\begin{thebibliography}{}
\expandafter\ifx\csname natexlab\endcsname\relax\def\natexlab#1{#1}\fi
\providecommand{\url}[1]{\href{#1}{#1}}

\bibitem[{{Alvarez} {et~al.}(2012){Alvarez}, {Finlator}, \&
  {Trenti}}]{Alvarez2012}
{Alvarez}, M.~A., {Finlator}, K., \& {Trenti}, M. 2012, \apjl, 759, L38

\bibitem[{{Astropy Collaboration} {et~al.}(2018){Astropy Collaboration},
  {Price-Whelan}, {Sip{\H{o}}cz}, {G{\"u}nther}, {Lim}, {Crawford}, {Conseil},
  {Shupe}, {Craig}, {Dencheva}, {Ginsburg}, {Vand erPlas}, {Bradley},
  {P{\'e}rez-Su{\'a}rez}, {de Val-Borro}, {Aldcroft}, {Cruz}, {Robitaille},
  {Tollerud}, {Ardelean}, {Babej}, {Bach}, {Bachetti}, {Bakanov}, {Bamford},
  {Barentsen}, {Barmby}, {Baumbach}, {Berry}, {Biscani}, {Boquien}, {Bostroem},
  {Bouma}, {Brammer}, {Bray}, {Breytenbach}, {Buddelmeijer}, {Burke},
  {Calderone}, {Cano Rodr{\'\i}guez}, {Cara}, {Cardoso}, {Cheedella}, {Copin},
  {Corrales}, {Crichton}, {D'Avella}, {Deil}, {Depagne}, {Dietrich}, {Donath},
  {Droettboom}, {Earl}, {Erben}, {Fabbro}, {Ferreira}, {Finethy}, {Fox},
  {Garrison}, {Gibbons}, {Goldstein}, {Gommers}, {Greco}, {Greenfield},
  {Groener}, {Grollier}, {Hagen}, {Hirst}, {Homeier}, {Horton}, {Hosseinzadeh},
  {Hu}, {Hunkeler}, {Ivezi{\'c}}, {Jain}, {Jenness}, {Kanarek}, {Kendrew},
  {Kern}, {Kerzendorf}, {Khvalko}, {King}, {Kirkby}, {Kulkarni}, {Kumar},
  {Lee}, {Lenz}, {Littlefair}, {Ma}, {Macleod}, {Mastropietro}, {McCully},
  {Montagnac}, {Morris}, {Mueller}, {Mumford}, {Muna}, {Murphy}, {Nelson},
  {Nguyen}, {Ninan}, {N{\"o}the}, {Ogaz}, {Oh}, {Parejko}, {Parley}, {Pascual},
  {Patil}, {Patil}, {Plunkett}, {Prochaska}, {Rastogi}, {Reddy Janga},
  {Sabater}, {Sakurikar}, {Seifert}, {Sherbert}, {Sherwood-Taylor}, {Shih},
  {Sick}, {Silbiger}, {Singanamalla}, {Singer}, {Sladen}, {Sooley},
  {Sornarajah}, {Streicher}, {Teuben}, {Thomas}, {Tremblay}, {Turner},
  {Terr{\'o}n}, {van Kerkwijk}, {de la Vega}, {Watkins}, {Weaver}, {Whitmore},
  {Woillez}, {Zabalza}, \& {Astropy Contributors}}]{astropy:2018}
{Astropy Collaboration}, {Price-Whelan}, A.~M., {Sip{\H{o}}cz}, B.~M., {et~al.}
  2018, \aj, 156, 123

\bibitem[{{Atek} {et~al.}(2018){Atek}, {Richard}, {Kneib}, \&
  {Schaerer}}]{Atek2018}
{Atek}, H., {Richard}, J., {Kneib}, J.-P., \& {Schaerer}, D. 2018, \mnras, 479,
  5184

\bibitem[{{Ba{\~n}ados} {et~al.}(2018){Ba{\~n}ados}, {Venemans},
  {Mazzucchelli}, {Farina}, {Walter}, {Wang}, {Decarli}, {Stern}, {Fan},
  {Davies}, {Hennawi}, {Simcoe}, {Turner}, {Rix}, {Yang}, {Kelson}, {Rudie}, \&
  {Winters}}]{Banados2018}
{Ba{\~n}ados}, E., {Venemans}, B.~P., {Mazzucchelli}, C., {et~al.} 2018, \nat,
  553, 473

\bibitem[{{Behrens} {et~al.}(2018){Behrens}, {Pallottini}, {Ferrara},
  {Gallerani}, \& {Vallini}}]{Behrens2018}
{Behrens}, C., {Pallottini}, A., {Ferrara}, A., {Gallerani}, S., \& {Vallini},
  L. 2018, \mnras, 477, 552

\bibitem[{{Bertin} \& {Arnouts}(1996)}]{Bertin1996}
{Bertin}, E., \& {Arnouts}, S. 1996, \aaps, 117, 393

\bibitem[{{Bouwens} {et~al.}(2015){Bouwens}, {Illingworth}, {Oesch}, {Trenti},
  {Labb{\'e}}, {Bradley}, {Carollo}, {van Dokkum}, {Gonzalez}, {Holwerda},
  {Franx}, {Spitler}, {Smit}, \& {Magee}}]{Bouwens2015}
{Bouwens}, R.~J., {Illingworth}, G.~D., {Oesch}, P.~A., {et~al.} 2015, \apj,
  803, 34

\bibitem[{{Bowler} {et~al.}(2020){Bowler}, {Jarvis}, {Dunlop}, {McLure},
  {McLeod}, {Adams}, {Milvang-Jensen}, \& {McCracken}}]{Bowler2020}
{Bowler}, R.~A.~A., {Jarvis}, M.~J., {Dunlop}, J.~S., {et~al.} 2020, \mnras,
  493, 2059

\bibitem[{{Brada{\v{c}}} {et~al.}(2019){Brada{\v{c}}}, {Huang}, {Fontana},
  {Castellano}, {Merlin}, {Amor{\'\i}n}, {Hoag}, {Strait}, {Santini}, {Ryan},
  {Casertano}, {Lemaux}, {Lubin}, {Schmidt}, {Schrabback}, {Treu}, {von der
  Linden}, {Mason}, \& {Wang}}]{Bradac2019}
{Brada{\v{c}}}, M., {Huang}, K.-H., {Fontana}, A., {et~al.} 2019, \mnras, 489,
  99

\bibitem[{{Brammer} {et~al.}(2008){Brammer}, {van Dokkum}, \&
  {Coppi}}]{Brammer2008}
{Brammer}, G.~B., {van Dokkum}, P.~G., \& {Coppi}, P. 2008, \apj, 686, 1503

\bibitem[{{Bruzual} \& {Charlot}(2003)}]{BC03}
{Bruzual}, G., \& {Charlot}, S. 2003, \mnras, 344, 1000

\bibitem[{{Calzetti} {et~al.}(2000){Calzetti}, {Armus}, {Bohlin}, {Kinney},
  {Koornneef}, \& {Storchi-Bergmann}}]{Calzetti2000}
{Calzetti}, D., {Armus}, L., {Bohlin}, R.~C., {et~al.} 2000, \apj, 533, 682

\bibitem[{{Carnall} {et~al.}(2019){Carnall}, {Leja}, {Johnson}, {McLure},
  {Dunlop}, \& {Conroy}}]{Carnall2019}
{Carnall}, A.~C., {Leja}, J., {Johnson}, B.~D., {et~al.} 2019, \apj, 873, 44

\bibitem[{{Carnall} {et~al.}(2018){Carnall}, {McLure}, {Dunlop}, \&
  {Dav{\'e}}}]{Carnall2018}
{Carnall}, A.~C., {McLure}, R.~J., {Dunlop}, J.~S., \& {Dav{\'e}}, R. 2018,
  \mnras, 480, 4379

\bibitem[{{Castellano} {et~al.}(2016){Castellano}, {Amor{\'\i}n}, {Merlin},
  {Fontana}, {McLure}, {M{\'a}rmol-Queralt{\'o}}, {Mortlock}, {Parsa},
  {Dunlop}, {Elbaz}, {Balestra}, {Boucaud}, {Bourne}, {Boutsia}, {Brammer},
  {Bruce}, {Buitrago}, {Capak}, {Cappelluti}, {Ciesla}, {Comastri}, {Cullen},
  {Derriere}, {Faber}, {Giallongo}, {Grazian}, {Grillo}, {Mercurio},
  {Micha{\l}owski}, {Nonino}, {Paris}, {Pentericci}, {Pilo}, {Rosati},
  {Santini}, {Schreiber}, {Shu}, \& {Wang}}]{Castellano2016}
{Castellano}, M., {Amor{\'\i}n}, R., {Merlin}, E., {et~al.} 2016, \aap, 590,
  A31

\bibitem[{{Castellano} {et~al.}(2022){Castellano}, {Fontana}, {Treu},
  {Santini}, {Merlin}, {Leethochawalit}, {Trenti}, {Mestric}, {Vanzella},
  {Bonchi}, {Belfiori}, {Nonino}, {Paris}, {Polenta}, {Roberts-Borsani},
  {Boyett}, {Calabro}, {Glazebrook}, {Grillo}, {Mascia}, {Mason}, {Mercurio},
  {Morishita}, {Nanayakkara}, {Pentericci}, {Rosati}, {Vulcani}, {Wang}, \&
  {Yang}}]{Castellano2022}
{Castellano}, M., {Fontana}, A., {Treu}, T., {et~al.} 2022, arXiv e-prints,
  arXiv:2207.09436

\bibitem[{{Diemer} {et~al.}(2017){Diemer}, {Sparre}, {Abramson}, \&
  {Torrey}}]{Diemer2017}
{Diemer}, B., {Sparre}, M., {Abramson}, L.~E., \& {Torrey}, P. 2017, \apj, 839,
  26

\bibitem[{{Ferland} {et~al.}(2017){Ferland}, {Chatzikos}, {Guzm{\'a}n},
  {Lykins}, {van Hoof}, {Williams}, {Abel}, {Badnell}, {Keenan}, {Porter}, \&
  {Stancil}}]{Ferland2017}
{Ferland}, G.~J., {Chatzikos}, M., {Guzm{\'a}n}, F., {et~al.} 2017, \rmxaa, 53,
  385

\bibitem[{{Fontana} {et~al.}(2000){Fontana}, {D'Odorico}, {Poli}, {Giallongo},
  {Arnouts}, {Cristiani}, {Moorwood}, \& {Saracco}}]{Fontana2000}
{Fontana}, A., {D'Odorico}, S., {Poli}, F., {et~al.} 2000, \aj, 120, 2206

\bibitem[{{Franco} {et~al.}(2018){Franco}, {Elbaz}, {B{\'e}thermin},
  {Magnelli}, {Schreiber}, {Ciesla}, {Dickinson}, {Nagar}, {Silverman},
  {Daddi}, {Alexander}, {Wang}, {Pannella}, {Le Floc'h}, {Pope}, {Giavalisco},
  {Maury}, {Bournaud}, {Chary}, {Demarco}, {Ferguson}, {Finkelstein}, {Inami},
  {Iono}, {Juneau}, {Lagache}, {Leiton}, {Lin}, {Magdis}, {Messias},
  {Motohara}, {Mullaney}, {Okumura}, {Papovich}, {Pforr}, {Rujopakarn},
  {Sargent}, {Shu}, \& {Zhou}}]{Franco2018}
{Franco}, M., {Elbaz}, D., {B{\'e}thermin}, M., {et~al.} 2018, \aap, 620, A152

\bibitem[{{Gladders} {et~al.}(2013){Gladders}, {Oemler}, {Dressler},
  {Poggianti}, {Vulcani}, \& {Abramson}}]{Gladders2013}
{Gladders}, M.~D., {Oemler}, A., {Dressler}, A., {et~al.} 2013, \apj, 770, 64

\bibitem[{{Glazebrook} {et~al.}(2022){Glazebrook}, {Nanayakkara}, {Jacobs},
  {Leethochawalit}, {Calabr{\`o}}, {Bonchi}, {Castellano}, {Fontana}, {Mason},
  {Merlin}, {Morishita}, {Paris}, {Trenti}, {Treu}, {Santini}, {Wang},
  {Boyett}, {Bradac}, {Brammer}, {Jones}, {Marchesini}, {Nonino}, \&
  {Vulcani}}]{Glazebrook2022}
{Glazebrook}, K., {Nanayakkara}, T., {Jacobs}, C., {et~al.} 2022, arXiv
  e-prints, arXiv:2208.03468

\bibitem[{{Greig} {et~al.}(2017){Greig}, {Mesinger}, {Haiman}, \&
  {Simcoe}}]{Greig2017}
{Greig}, B., {Mesinger}, A., {Haiman}, Z., \& {Simcoe}, R.~A. 2017, \mnras,
  466, 4239

\bibitem[{{Hainline} {et~al.}(2020){Hainline}, {Hviding}, {Rieke}, {Shivaei},
  {Endsley}, {Curtis-Lake}, {Smit}, {Williams}, {Alberts}, {K Boyett},
  {Bunker}, {Egami}, {Maseda}, {Tacchella}, \& {Willmer}}]{Hainline2020}
{Hainline}, K.~N., {Hviding}, R.~E., {Rieke}, M., {et~al.} 2020, \apj, 892, 125

\bibitem[{{Inoue} {et~al.}(2014){Inoue}, {Shimizu}, {Iwata}, \&
  {Tanaka}}]{Inoue2014}
{Inoue}, A.~K., {Shimizu}, I., {Iwata}, I., \& {Tanaka}, M. 2014, \mnras, 442,
  1805

\bibitem[{{Kauffmann} {et~al.}(2020){Kauffmann}, {Le F{\`e}vre}, {Ilbert},
  {Chevallard}, {Williams}, {Curtis-Lake}, {Colina}, {P{\'e}rez-Gonz{\'a}lez},
  {Pye}, \& {Caputi}}]{Kauffmann2020}
{Kauffmann}, O.~B., {Le F{\`e}vre}, O., {Ilbert}, O., {et~al.} 2020, \aap, 640,
  A67

\bibitem[{{Kroupa} \& {Boily}(2002)}]{Kroupa2002}
{Kroupa}, P., \& {Boily}, C.~M. 2002, \mnras, 336, 1188

\bibitem[{{Labb{\'e}} {et~al.}(2010){Labb{\'e}}, {Gonz{\'a}lez}, {Bouwens},
  {Illingworth}, {Franx}, {Trenti}, {Oesch}, {van Dokkum}, {Stiavelli},
  {Carollo}, {Kriek}, \& {Magee}}]{Labbe2010}
{Labb{\'e}}, I., {Gonz{\'a}lez}, V., {Bouwens}, R.~J., {et~al.} 2010, \apjl,
  716, L103

\bibitem[{{Labb{\'e}} {et~al.}(2015){Labb{\'e}}, {Oesch}, {Illingworth}, {van
  Dokkum}, {Bouwens}, {Franx}, {Carollo}, {Trenti}, {Holden}, {Smit},
  {Gonz{\'a}lez}, {Magee}, {Stiavelli}, \& {Stefanon}}]{Labbe2015ApJS}
{Labb{\'e}}, I., {Oesch}, P.~A., {Illingworth}, G.~D., {et~al.} 2015, \apjs,
  221, 23

\bibitem[{{Laporte} {et~al.}(2017){Laporte}, {Ellis}, {Boone}, {Bauer},
  {Qu{\'e}nard}, {Roberts-Borsani}, {Pell{\'o}}, {P{\'e}rez-Fournon}, \&
  {Streblyanska}}]{Laporte2017}
{Laporte}, N., {Ellis}, R.~S., {Boone}, F., {et~al.} 2017, \apjl, 837, L21

\bibitem[{{Leethochawalit} {et~al.}(2019){Leethochawalit}, {Kirby}, {Ellis},
  {Moran}, \& {Treu}}]{Leethochawalit2019}
{Leethochawalit}, N., {Kirby}, E.~N., {Ellis}, R.~S., {Moran}, S.~M., \&
  {Treu}, T. 2019, \apj, 885, 100

\bibitem[{{Leja} {et~al.}(2019){Leja}, {Carnall}, {Johnson}, {Conroy}, \&
  {Speagle}}]{Leja2019}
{Leja}, J., {Carnall}, A.~C., {Johnson}, B.~D., {Conroy}, C., \& {Speagle},
  J.~S. 2019, \apj, 876, 3

\bibitem[{{Ma} {et~al.}(2018){Ma}, {Hopkins}, {Garrison-Kimmel},
  {Faucher-Gigu{\`e}re}, {Quataert}, {Boylan-Kolchin}, {Hayward}, {Feldmann},
  \& {Kere{\v{s}}}}]{Ma2018}
{Ma}, X., {Hopkins}, P.~F., {Garrison-Kimmel}, S., {et~al.} 2018, \mnras, 478,
  1694

\bibitem[{{Maraston} {et~al.}(2010){Maraston}, {Pforr}, {Renzini}, {Daddi},
  {Dickinson}, {Cimatti}, \& {Tonini}}]{Maraston2010}
{Maraston}, C., {Pforr}, J., {Renzini}, A., {et~al.} 2010, \mnras, 407, 830

\bibitem[{{Mason} {et~al.}(2018){Mason}, {Treu}, {Dijkstra}, {Mesinger},
  {Trenti}, {Pentericci}, {de Barros}, \& {Vanzella}}]{Mason2018}
{Mason}, C.~A., {Treu}, T., {Dijkstra}, M., {et~al.} 2018, \apj, 856, 2

\bibitem[{{McGreer} {et~al.}(2015){McGreer}, {Mesinger}, \&
  {D'Odorico}}]{McGreer2015}
{McGreer}, I.~D., {Mesinger}, A., \& {D'Odorico}, V. 2015, \mnras, 447, 499

\bibitem[{{Medezinski} {et~al.}(2016){Medezinski}, {Umetsu}, {Okabe}, {Nonino},
  {Molnar}, {Massey}, {Dupke}, \& {Merten}}]{Medezinski2016}
{Medezinski}, E., {Umetsu}, K., {Okabe}, N., {et~al.} 2016, \apj, 817, 24

\bibitem[{{Merlin} {et~al.}(2019){Merlin}, {Pilo}, {Fontana}, {Castellano},
  {Paris}, {Roscani}, {Santini}, \& {Torelli}}]{Merlin2019}
{Merlin}, E., {Pilo}, S., {Fontana}, A., {et~al.} 2019, \aap, 622, A169

\bibitem[{{Merlin} {et~al.}(2015){Merlin}, {Fontana}, {Ferguson}, {Dunlop},
  {Elbaz}, {Bourne}, {Bruce}, {Buitrago}, {Castellano}, {Schreiber}, {Grazian},
  {McLure}, {Okumura}, {Shu}, {Wang}, {Amor{\'\i}n}, {Boutsia}, {Cappelluti},
  {Comastri}, {Derriere}, {Faber}, \& {Santini}}]{Merlin2015}
{Merlin}, E., {Fontana}, A., {Ferguson}, H.~C., {et~al.} 2015, \aap, 582, A15

\bibitem[{{Merlin} {et~al.}(2022){Merlin}, {Bonchi}, {Paris}, {Belfiori},
  {Fontana}, {Castellano}, {Nonino}, {Polenta}, {Santini}, {Yang},
  {Glazebrook}, {Treu}, {Roberts-Borsani}, {Trenti}, {Birrer}, {Brammer},
  {Grillo}, {Calabr{\`o}}, {Marchesini}, {Mason}, {Mercurio}, {Morishita},
  {Strait}, {Boyett}, {Leethochawalit}, {Nanayakkara}, {Vulcani}, {Bradac}, \&
  {Wang}}]{Merlin2022}
{Merlin}, E., {Bonchi}, A., {Paris}, D., {et~al.} 2022, arXiv e-prints,
  arXiv:2207.11701

\bibitem[{{Morishita} {et~al.}(2020){Morishita}, {Stiavelli}, {Trenti}, {Treu},
  {Roberts-Borsani}, {Mason}, {Hashimoto}, {Bradley}, {Coe}, \&
  {Ishikawa}}]{Morishita2020}
{Morishita}, T., {Stiavelli}, M., {Trenti}, M., {et~al.} 2020, \apj, 904, 50

\bibitem[{{Naidu} {et~al.}(2022){Naidu}, {Oesch}, {van Dokkum}, {Nelson},
  {Suess}, {Whitaker}, {Allen}, {Bezanson}, {Bouwens}, {Brammer}, {Conroy},
  {Illingworth}, {Labbe}, {Leja}, {Leonova}, {Matthee}, {Price}, {Setton},
  {Strait}, {Stefanon}, {Tacchella}, {Toft}, {Weaver}, \& {Weibel}}]{Naidu2022}
{Naidu}, R.~P., {Oesch}, P.~A., {van Dokkum}, P., {et~al.} 2022, arXiv
  e-prints, arXiv:2207.09434

\bibitem[{{Nanayakkara} {et~al.}(2022){Nanayakkara}, {Glazebrook}, {Jacobs},
  {Bonchi}, {Castellano}, {Fontana}, {Mason}, {Merlin}, {Morishita}, {Paris},
  {Trenti}, {Treu}, {Calabro}, {Boyett}, {Bradac}, {Leethochawalit},
  {Marchesini}, {Santini}, {Strait}, {Vanzella}, {Vulcani}, {Wang}, \&
  {Yang}}]{Nanayakkara2022}
{Nanayakkara}, T., {Glazebrook}, K., {Jacobs}, C., {et~al.} 2022, arXiv
  e-prints, arXiv:2207.13860

\bibitem[{{Oke} \& {Gunn}(1983)}]{Oke1983}
{Oke}, J.~B., \& {Gunn}, J.~E. 1983, \apj, 266, 713

\bibitem[{{Rigby} {et~al.}(2022){Rigby}, {Perrin}, {McElwain}, {Kimble},
  {Friedman}, {Lallo}, {Doyon}, {Feinberg}, {Ferruit}, {Glasse}, {Rieke},
  {Rieke}, {Wright}, {Willott}, {Colon}, {Milam}, {Neff}, {Stark}, {Valenti},
  {Abell}, {Abney}, {Abul-Huda}, {Acton}, {Adams}, {Adler}, {Aguilar}, {Ahmed},
  {Albert}, {Alberts}, {Aldridge}, {Allen}, {Altenburg}, {Alvarez Marquez},
  {Alves de Oliveira}, {Andersen}, {Anderson}, {Anderson}, {Argyriou},
  {Armstrong}, {Arribas}, {Artigau}, {Arvai}, {Atkinson}, {Bacon}, {Bair},
  {Banks}, {Barrientes}, {Barringer}, {Bartosik}, {Bast}, {Baudoz}, {Beatty},
  {Bechtold}, {Beck}, {Bergeron}, {Bergkoetter}, {Bhatawdekar}, {Birkmann},
  {Blazek}, {Blome}, {Boccaletti}, {Boeker}, {Boia}, {Bonaventura}, {Bond},
  {Bosley}, {Boucarut}, {Bourque}, {Bouwman}, {Bower}, {Bowers}, {Boyer},
  {Bradley}, {Brady}, {Braun}, {Breda}, {Bresnahan}, {Bright}, {Britt},
  {Bromenschenkel}, {Brooks}, {Brooks}, {Brown}, {Brown}, {Brown}, {Bunker},
  {Burger}, {Bushouse}, {Cale}, {Cameron}, {Cameron}, {Canipe}, {Caplinger},
  {Caputo}, {Cara}, {Carey}, {Carniani}, {Carrasquilla}, {Carruthers}, {Case},
  {Catherine}, {Chance}, {Chapman}, {Charlot}, {Charlow}, {Chayer}, {Chen},
  {Cherinka}, {Chichester}, {Chilton}, {Chonis}, {Clampin}, {Clark}, {Clark},
  {Coe}, {Coleman}, {Comber}, {Comeau}, {Connolly}, {Cooper}, {Cooper},
  {Coppock}, {Correnti}, {Cossou}, {Coulais}, {Coyle}, {Cracraft}, {Curti},
  {Cuturic}, {Davis}, {Davis}, {Dean}, {DeLisa}, {deMeester}, {Dencheva},
  {Dencheva}, {DePasquale}, {Deschenes}, {Hunor Detre}, {Diaz}, {Dicken},
  {DiFelice}, {Dillman}, {Dixon}, {Doggett}, {Donaldson}, {Douglas}, {DuPrie},
  {Dupuis}, {Durning}, {Easmin}, {Eck}, {Edeani}, {Egami}, {Ehrenwinkler},
  {Eisenhamer}, {Eisenhower}, {Elie}, {Elliott}, {Elliott}, {Ellis},
  {Engesser}, {Espinoza}, {Etienne}, {Etxaluze}, {Falini}, {Feeney}, {Ferry},
  {Filippazzo}, {Fincham}, {Fix}, {Flagey}, {Florian}, {Flynn}, {Fontanella},
  {Ford}, {Forshay}, {Fox}, {Franz}, {Fu}, {Fullerton}, {Galkin}, {Galyer},
  {Garcia Marin}, {Gardner}, {Gardner}, {Garland}, {Garrett}, {Gasman},
  {Gaspar}, {Gaudreau}, {Gauthier}, {Geers}, {Geithner}, {Gennaro}, {Giardino},
  {Girard}, {Giuliano}, {Glassmire}, {Glauser}, {Glazer}, {Godfrey},
  {Golimowski}, {Gollnitz}, {Gong}, {Gonzaga}, {Gordon}, {Gordon},
  {Goudfrooij}, {Greene}, {Greenhouse}, {Grimaldi}, {Groebner}, {Grundy},
  {Guillard}, {Gutman}, {Ha}, {Haderlein}, {Hagedorn}, {Hainline}, {Haley},
  {Hami}, {Hamilton}, {Hammel}, {Hansen}, {Harkins}, {Harr}, {Hart}, {Hart},
  {Hartig}, {Hashimoto}, {Haskins}, {Hathaway}, {Havey}, {Hayden}, {Hecht},
  {Heller-Boyer}, {Henriques}, {Henry}, {Hermann}, {Hernandez}, {Hesman},
  {Hicks}, {Hilbert}, {Hines}, {Hoffman}, {Holfeltz}, {Holler}, {Hoppa},
  {Hott}, {Howard}, {Howard}, {Hunter}, {Hunter}, {Hurst}, {Husemann},
  {Hustak}, {Ilinca Ignat}, {Illingworth}, {Irish}, {Jackson}, {Jahromi},
  {Jakobsen}, {James}, {James}, {Januszewski}, {Jenkins}, {Jirdeh}, {Johnson},
  {Johnson}, {Jones}, {Jones}, {Jones}, {Jones}, {Jordan}, {Jordan}, {Jurczyk},
  {Jurling}, {Kaleida}, {Kalmanson}, {Kammerer}, {Kang}, {Kao}, {Karakla},
  {Kavanagh}, {Kelly}, {Kendrew}, {Kennedy}, {Kenny}, {Keski-kuha}, {Keyes},
  {Kidwell}, {Kinzel}, {Kirk}, {Kirkpatrick}, {Kirshenblat}, {Klaassen},
  {Knapp}, {Knight}, {Knollenberg}, {Koehler}, {Koekemoer}, {Kovacs}, {Kulp},
  {Kumari}, {Kyprianou}, {La Massa}, {Labador}, {Labiano Ortega}, {Lagage},
  {Lajoie}, {Lallo}, {Lam}, {Lamb}, {Lambros}, {Lampenfield}, {Langston},
  {Larson}, {Law}, {Lawrence}, {Lee}, {Leisenring}, {Lepo}, {Leveille},
  {Levenson}, {Levine}, {Levy}, {Lewis}, {Lewis}, {Libralato}, {Lightsey},
  {Link}, {Liu}, {Lo}, {Lockwood}, {Logue}, {Long}, {Long}, {Loomis},
  {Lopez-Caniego}, {Alvarez}, {Love-Pruitt}, {Lucy}, {Luetzgendorf}, {Maghami},
  {Maiolino}, {Major}, {Malla}, {Malumuth}, {Manjavacas}, {Mannfolk},
  {Marrione}, {Marston}, {Martel}, {Maschmann}, {Masci}, {Masciarelli},
  {Maszkiewicz}, {Mather}, {McKenzie}, {McLean}, {McMaster}, {Melbourne},
  {Mel{\'e}ndez}, {Menzel}, {Merz}, {Meyett}, {Meza}, {Miskey}, {Misselt},
  {Moller}, {Morrison}, {Morse}, {Moseley}, {Mosier}, {Mountain}, {Mueckay},
  {Mueller}, {Mullally}, {Murphy}, {Murray}, {Murray}, {Mustelier},
  {Muzerolle}, {Mycroft}, {Myers}, {Myrick}, {Nanavati}, {Nance}, {Nayak},
  {Naylor}, {Nelan}, {Nickson}, {Nielson}, {Nieto-Santisteban}, {Nikolov},
  {Noriega-Crespo}, {O'Shaughnessy}, {O'Sullivan}, {Ochs}, {Ogle}, {Oleszczuk},
  {Olmsted}, {Osborne}, {Ottens}, {Owens}, {Pacifici}, {Pagan}, {Page}, {Park},
  {Parrish}, {Patapis}, {Paul}, {Pauly}, {Pavlovsky}, {Pedder}, {Peek},
  {Pena-Guerrero}, {Pennanen}, {Perez}, {Perna}, {Perriello}, {Phillips},
  {Pietraszkiewicz}, {Pinaud}, {Pirzkal}, {Pitman}, {Piwowar}, {Platais},
  {Player}, {Plesha}, {Pollizi}, {Polster}, {Pontoppidan}, {Porterfield},
  {Proffitt}, {Pueyo}, {Pulliam}, {Quirt}, {Quispe Neira}, {Ramos Alarcon},
  {Ramsay}, {Rapp}, {Rapp}, {Rauscher}, {Ravindranath}, {Rawle}, {Regan},
  {Reichard}, {Reis}, {Ressler}, {Rest}, {Reynolds}, {Rhue}, {Richon},
  {Rickman}, {Ridgaway}, {Ritchie}, {Rix}, {Robberto}, {Robinson}, {Robinson},
  {Robinson}, {Rock}, {Rodriguez}, {Rodriguez Del Pino}, {Roellig}, {Rohrbach},
  {Roman}, {Romelfanger}, {Rose}, {Roteliuk}, {Roth}, {Rothwell}, {Rowlands},
  {Roy}, {Royer}, {Royle}, {Rui}, {Rumler}, {Runnels}, {Russ}, {Rustamkulov},
  {Ryden}, {Ryer}, {Sabata}, {Sabatke}, {Sabbi}, {Samuelson}, {Sapp},
  {Sappington}, {Sargent}, {Sauer}, {Scheithauer}, {Schlawin}, {Schlitz},
  {Schmitz}, {Schneider}, {Schreiber}, {Schulze}, {Schwab}, {Scott}, {Sembach},
  {Shanahan}, {Shaughnessy}, {Shaw}, {Shawger}, {Shay}, {Sheehan}, {Shen},
  {Sherman}, {Shiao}, {Shih}, {Shivaei}, {Sienkiewicz}, {Sing}, {Sirianni},
  {Sivaramakrishnan}, {Skipper}, {Sloan}, {Slocum}, {Slowinski}, {Smith},
  {Smith}, {Smith}, {Smith}, {Snyder}, {Soh}, {Sohn}, {Soto}, {Spencer},
  {Stallcup}, {Stansberry}, {Starr}, {Starr}, {Stewart}, {Stiavelli},
  {Straughn}, {Strickland}, {Stys}, {Summers}, {Sun}, {Sunnquist}, {Swade},
  {Swam}, {Swaters}, {Swoish}, {Taylor}, {Taylor}, {Te Plate}, {Tea}, {Teague},
  {Telfer}, {Temim}, {Thatte}, {Thompson}, {Thompson}, {Thomson}, {Tikkanen},
  {Tippet}, {Todd}, {Toolan}, {Tran}, {Trejo}, {Truong}, {Tsukamoto},
  {Tustain}, {Tyra}, {Ubeda}, {Underwood}, {Uzzo}, {Van Campen}, {Vandal},
  {Vandenbussche}, {Vila}, {Volk}, {Wahlgren}, {Waldman}, {Walker}, {Wander},
  {Warfield}, {Warner}, {Wasiak}, {Watkins}, {Weaver}, {Weilert}, {Weiser},
  {Weiss}, {Weissman}, {Welty}, {West}, {Wheate}, {Wheatley}, {Wheeler},
  {White}, {Whiteaker}, {Whitehouse}, {Whiteleather}, {Whitman}, {Williams},
  {Willmer}, {Willoughby}, {Wilson}, {Wirth}, {Wislowski}, {Wolf}, {Wolfe},
  {Wolff}, {Workman}, {Wright}, {Wu}, {Wu}, {Wymer}, {Yates}, {Yeager},
  {Yeates}, {Yerger}, {Yoon}, {Young}, {Yu}, {Zak}, {Zeidler}, {Zhou},
  {Zielinski}, {Zincke}, \& {Zonak}}]{Rigby2022}
{Rigby}, J., {Perrin}, M., {McElwain}, M., {et~al.} 2022, arXiv e-prints,
  arXiv:2207.05632

\bibitem[{{Roberts-Borsani} {et~al.}(2022){Roberts-Borsani}, {Morishita},
  {Treu}, {Leethochawalit}, \& {Trenti}}]{Roberts-Borsani2022}
{Roberts-Borsani}, G., {Morishita}, T., {Treu}, T., {Leethochawalit}, N., \&
  {Trenti}, M. 2022, \apj, 927, 236

\bibitem[{{Roberts-Borsani} {et~al.}(2021){Roberts-Borsani}, {Treu}, {Mason},
  {Schmidt}, {Jones}, \& {Fontana}}]{rb21}
{Roberts-Borsani}, G., {Treu}, T., {Mason}, C., {et~al.} 2021, \apj, 910, 86

\bibitem[{{Roberts-Borsani} {et~al.}(2016){Roberts-Borsani}, {Bouwens},
  {Oesch}, {Labbe}, {Smit}, {Illingworth}, {van Dokkum}, {Holden}, {Gonzalez},
  {Stefanon}, {Holwerda}, \& {Wilkins}}]{Roberts-Borsani2016}
{Roberts-Borsani}, G.~W., {Bouwens}, R.~J., {Oesch}, P.~A., {et~al.} 2016,
  \apj, 823, 143

\bibitem[{{Robertson} {et~al.}(2015){Robertson}, {Ellis}, {Furlanetto}, \&
  {Dunlop}}]{Robertson2015}
{Robertson}, B.~E., {Ellis}, R.~S., {Furlanetto}, S.~R., \& {Dunlop}, J.~S.
  2015, \apjl, 802, L19

\bibitem[{{Ryan} \& {Reid}(2016)}]{Ryan_Jr2016}
{Ryan}, R.~E., J., \& {Reid}, I.~N. 2016, \aj, 151, 92

\bibitem[{{Santini} {et~al.}(2015){Santini}, {Ferguson}, {Fontana}, {Mobasher},
  {Barro}, {Castellano}, {Finkelstein}, {Grazian}, {Hsu}, {Lee}, {Lee},
  {Pforr}, {Salvato}, {Wiklind}, {Wuyts}, {Almaini}, {Cooper}, {Galametz},
  {Weiner}, {Amorin}, {Boutsia}, {Conselice}, {Dahlen}, {Dickinson},
  {Giavalisco}, {Grogin}, {Guo}, {Hathi}, {Kocevski}, {Koekemoer},
  {Kurczynski}, {Merlin}, {Mortlock}, {Newman}, {Paris}, {Pentericci},
  {Simons}, \& {Willner}}]{Santini2015}
{Santini}, P., {Ferguson}, H.~C., {Fontana}, A., {et~al.} 2015, \apj, 801, 97

\bibitem[{{Santini} {et~al.}(2017){Santini}, {Fontana}, {Castellano}, {Di
  Criscienzo}, {Merlin}, {Amorin}, {Cullen}, {Daddi}, {Dickinson}, {Dunlop},
  {Grazian}, {Lamastra}, {McLure}, {Micha{\l}owski}, {Pentericci}, \&
  {Shu}}]{Santini2017}
{Santini}, P., {Fontana}, A., {Castellano}, M., {et~al.} 2017, \apj, 847, 76

\bibitem[{{Santini} {et~al.}(2022){Santini}, {Fontana}, {Castellano},
  {Leethochawalit}, {Trenti}, {Treu}, {Belfiori}, {Birrer}, {Bonchi}, {Merlin},
  {Mason}, {Morishita}, {Nonino}, {Paris}, {Polenta}, {Rosati}, {Yang},
  {Bradac}, {Calabr{\`o}}, {Dressler}, {Glazebrook}, {Marchesini}, {Mascia},
  {Nanayakkara}, {Pentericci}, {Roberts-Borsani}, {Scarlata}, {Vulcani}, \&
  {Wang}}]{Santini2022}
---. 2022, arXiv e-prints, arXiv:2207.11379

\bibitem[{{Shu} {et~al.}(2022){Shu}, {Yang}, {Liu}, {Wang}, {Wang}, {Han},
  {Huang}, {Lim}, {Chang}, {Zheng}, {Zheng}, {Wang}, \& {Kong}}]{Shu2022}
{Shu}, X., {Yang}, L., {Liu}, D., {et~al.} 2022, \apj, 926, 155

\bibitem[{{Skelton} {et~al.}(2014){Skelton}, {Whitaker}, {Momcheva}, {Brammer},
  {van Dokkum}, {Labb{\'e}}, {Franx}, {van der Wel}, {Bezanson}, {Da Cunha},
  {Fumagalli}, {F{\"o}rster Schreiber}, {Kriek}, {Leja}, {Lundgren}, {Magee},
  {Marchesini}, {Maseda}, {Nelson}, {Oesch}, {Pacifici}, {Patel}, {Price},
  {Rix}, {Tal}, {Wake}, \& {Wuyts}}]{Skelton2014}
{Skelton}, R.~E., {Whitaker}, K.~E., {Momcheva}, I.~G., {et~al.} 2014, \apjs,
  214, 24

\bibitem[{{Stefanon} {et~al.}(2022{\natexlab{a}}){Stefanon}, {Bouwens},
  {Illingworth}, {Labb{\'e}}, {Oesch}, \& {Gonzalez}}]{Stefanon2022b}
{Stefanon}, M., {Bouwens}, R.~J., {Illingworth}, G.~D., {et~al.}
  2022{\natexlab{a}}, arXiv e-prints, arXiv:2204.02986

\bibitem[{{Stefanon} {et~al.}(2022{\natexlab{b}}){Stefanon}, {Bouwens},
  {Labb{\'e}}, {Illingworth}, {Oesch}, {van Dokkum}, \&
  {Gonzalez}}]{Stefanon2022a_stack}
{Stefanon}, M., {Bouwens}, R.~J., {Labb{\'e}}, I., {et~al.} 2022{\natexlab{b}},
  \apj, 927, 48

\bibitem[{{Steidel} {et~al.}(2016){Steidel}, {Strom}, {Pettini}, {Rudie},
  {Reddy}, \& {Trainor}}]{Steidel2016}
{Steidel}, C.~C., {Strom}, A.~L., {Pettini}, M., {et~al.} 2016, \apj, 826, 159

\bibitem[{{Straatman} {et~al.}(2016){Straatman}, {Spitler}, {Quadri},
  {Labb{\'e}}, {Glazebrook}, {Persson}, {Papovich}, {Tran}, {Brammer},
  {Cowley}, {Tomczak}, {Nanayakkara}, {Alcorn}, {Allen}, {Broussard}, {van
  Dokkum}, {Forrest}, {van Houdt}, {Kacprzak}, {Kawinwanichakij}, {Kelson},
  {Lee}, {McCarthy}, {Mehrtens}, {Monson}, {Murphy}, {Rees}, {Tilvi}, \&
  {Whitaker}}]{Straatman16}
{Straatman}, C. M.~S., {Spitler}, L.~R., {Quadri}, R.~F., {et~al.} 2016, \apj,
  830, 51

\bibitem[{{Strait} {et~al.}(2021){Strait}, {Brada{\v{c}}}, {Coe}, {Lemaux},
  {Carnall}, {Bradley}, {Pelliccia}, {Sharon}, {Zitrin}, {Acebron}, {Neufeld},
  {Andrade-Santos}, {Avila}, {Frye}, {Mahler}, {Nonino}, {Ogaz}, {Oguri},
  {Ouchi}, {Paterno-Mahler}, {Stark}, {Mainali}, {Oesch}, {Trenti}, {Carrasco},
  {Dawson}, {Jones}, {Umetsu}, \& {Vulcani}}]{strait21}
{Strait}, V., {Brada{\v{c}}}, M., {Coe}, D., {et~al.} 2021, \apj, 910, 135

\bibitem[{{Tacchella} {et~al.}(2022){Tacchella}, {Finkelstein}, {Bagley},
  {Dickinson}, {Ferguson}, {Giavalisco}, {Graziani}, {Grogin}, {Hathi},
  {Hutchison}, {Jung}, {Koekemoer}, {Larson}, {Papovich}, {Pirzkal},
  {Rojas-Ruiz}, {Song}, {Schneider}, {Somerville}, {Wilkins}, \&
  {Yung}}]{Tacchella2022}
{Tacchella}, S., {Finkelstein}, S.~L., {Bagley}, M., {et~al.} 2022, \apj, 927,
  170

\bibitem[{{Theios} {et~al.}(2019){Theios}, {Steidel}, {Strom}, {Rudie},
  {Trainor}, \& {Reddy}}]{Theios2019}
{Theios}, R.~L., {Steidel}, C.~C., {Strom}, A.~L., {et~al.} 2019, \apj, 871,
  128

\bibitem[{{Topping} {et~al.}(2022){Topping}, {Stark}, {Endsley}, {Bouwens},
  {Schouws}, {Smit}, {Stefanon}, {Inami}, {Bowler}, {Oesch}, {Gonzalez},
  {Dayal}, {da Cunha}, {Algera}, {van der Werf}, {Pallottini}, {Barrufet De
  Soto}, {Schneider}, {De Looze}, {Sommovigo}, {Whitler}, {Graziani},
  {Fudamoto}, \& {Ferrara}}]{Topping2022}
{Topping}, M.~W., {Stark}, D.~P., {Endsley}, R., {et~al.} 2022, arXiv e-prints,
  arXiv:2203.07392

\bibitem[{{Treu} {et~al.}(2013){Treu}, {Schmidt}, {Trenti}, {Bradley}, \&
  {Stiavelli}}]{Treu2013}
{Treu}, T., {Schmidt}, K.~B., {Trenti}, M., {Bradley}, L.~D., \& {Stiavelli},
  M. 2013, \apjl, 775, L29

\bibitem[{{Treu} {et~al.}(2022{\natexlab{a}}){Treu}, {Roberts-Borsani},
  {Bradac}, {Brammer}, {Fontana}, {Henry}, {Mason}, {Morishita}, {Pentericci},
  {Wang}, {Acebron}, {Bagley}, {Bergamini}, {Belfiori}, {Bonchi}, {Boyett},
  {Boutsia}, {Calabro}, {Caminha}, {Castellano}, {Dressler}, {Glazebrook},
  {Grillo}, {Jacobs}, {Jones}, {Kelly}, {Leethochawalit}, {Malkan},
  {Marchesini}, {Mascia}, {Mercurio}, {Merlin}, {Nanayakkara}, {Paris},
  {Poggianti}, {Rosati}, {Santini}, {Scarlata}, {Shipley}, {Strait}, {Trenti},
  {Tubthong}, {Vanzella}, {Vulcani}, \& {Yang}}]{Treu2022}
{Treu}, T., {Roberts-Borsani}, G., {Bradac}, M., {et~al.} 2022{\natexlab{a}},
  arXiv e-prints, arXiv:2206.07978

\bibitem[{{Treu} {et~al.}(2022{\natexlab{b}}){Treu}, {Calabro}, {Castellano},
  {Leethochawalit}, {Merlin}, {Fontana}, {Yang}, {Morishita}, {Trenti},
  {Dressler}, {Mason}, {Paris}, {Pentericci}, {Roberts-Borsani}, {Vulcani},
  {Boyett}, {Bradac}, {Glazebrook}, {Jones}, {Marchesini}, {Mascia},
  {Nanayakkara}, {Santini}, {Strait}, {Vanzella}, \& {Wang}}]{Treu2022b}
{Treu}, T., {Calabro}, A., {Castellano}, M., {et~al.} 2022{\natexlab{b}}, arXiv
  e-prints, arXiv:2207.13527

\bibitem[{{Whitler} {et~al.}(2022){Whitler}, {Stark}, {Endsley}, {Leja},
  {Charlot}, \& {Chevallard}}]{Whitler2022}
{Whitler}, L., {Stark}, D.~P., {Endsley}, R., {et~al.} 2022, arXiv e-prints,
  arXiv:2206.05315

\bibitem[{{Williams} {et~al.}(2018){Williams}, {Curtis-Lake}, {Hainline},
  {Chevallard}, {Robertson}, {Charlot}, {Endsley}, {Stark}, {Willmer},
  {Alberts}, {Amorin}, {Arribas}, {Baum}, {Bunker}, {Carniani}, {Crandall},
  {Egami}, {Eisenstein}, {Ferruit}, {Husemann}, {Maseda}, {Maiolino}, {Rawle},
  {Rieke}, {Smit}, {Tacchella}, \& {Willott}}]{Williams2018}
{Williams}, C.~C., {Curtis-Lake}, E., {Hainline}, K.~N., {et~al.} 2018, \apjs,
  236, 33

\bibitem[{{Yang} {et~al.}(2022){Yang}, {Morishita}, {Leethochawalit},
  {Castellano}, {Calabro}, {Treu}, {Bonchi}, {Fontana}, {Mason}, {Merlin},
  {Paris}, {Trenti}, {Roberts-Borsani}, {Bradac}, {Vanzella}, {Vulcani},
  {Marchesini}, {Ding}, {Nanayakkara}, {Birrer}, {Glazebrook}, {Jones},
  {Boyett}, {Santini}, {Strait}, \& {Wang}}]{Yang2022}
{Yang}, L., {Morishita}, T., {Leethochawalit}, N., {et~al.} 2022, arXiv
  e-prints, arXiv:2207.13101

\bibitem[{{Yung} {et~al.}(2019){Yung}, {Somerville}, {Popping}, {Finkelstein},
  {Ferguson}, \& {Dav{\'e}}}]{Yung2019}
{Yung}, L.~Y.~A., {Somerville}, R.~S., {Popping}, G., {et~al.} 2019, \mnras,
  490, 2855

\end{thebibliography}
\bibliographystyle{aasjournal}



\end{document}